# Surface Split Decompositions and Subgraph Isomorphism in Graphs on Surfaces

Paul Bonsma*

**Abstract**

The Subgraph Isomorphism problem asks, given a *host graph* $G$ on $n$ vertices and a *pattern graph* $P$ on $k$ vertices, whether $G$ contains a subgraph isomorphic to $P$. The restriction of this problem to planar graphs has often been considered. After a sequence of improvements, the current best algorithm for planar graphs is a linear time algorithm by Dorn (STACS '10), with complexity $2^{O(k)} \cdot O(n)$.

We generalize this result, by giving an algorithm of the same complexity for graphs that can be embedded in surfaces of bounded genus. At the same time, we simplify the algorithm and analysis. The key to these improvements is the introduction of surface split decompositions for bounded genus graphs, which generalize sphere cut decompositions for planar graphs. We extend the algorithm for the problem of counting and generating all subgraphs isomorphic to $P$, even for the case where $P$ is disconnected. This answers an open question by Eppstein (SODA '95 / JGAA '99).

## 1 Introduction

The Subgraph Isomorphism problem asks, given a *host graph* $G$ on $n$ vertices and a *pattern graph* $P$ on $k$ vertices, whether $G$ contains a subgraph isomorphic to $P$. This is a well-studied problem that generalizes many other important problems, such as finding cliques, determining the girth, finding complete bipartite subgraphs, and finding a Hamilton path or cycle. See Eppstein [14] for a survey on previous results for this problem, and its many applications. This problem is NP-complete in general, even for planar graphs (it generalizes Hamilton Path). However, in many cases $P$ can be considered to be a small fixed graph. In that case, a trivial polynomial time algorithm of complexity $n^{O(k)}$ exists. For general graphs, nothing better is known. When restricting $G$ to be planar, this can be improved significantly: Eppstein [13, 14] gave a linear time algorithm for Planar Subgraph Isomorphism for any fixed graph $P$ on $k$ vertices. This seems best possible. However to judge the practicality of such an algorithm, the dependency of the complexity on the value $k$ is also essential. This is where the refined viewpoint of parameterized complexity proves to be useful [12, 19]. We choose $k$ to be the parameter. An algorithm for this problem is then *Fixed Parameter Tractable (FPT)* if its complexity can be bounded by $f(k) \cdot O(n^c)$, where $c$ is a constant independent of $k$, and $f(k)$ is an arbitrary computable function.

When viewed as a parameterized problem, the complexity of Eppstein's algorithm [14] is $2^{O(k \log k)} \cdot O(n)$, hence it is an FPT algorithm. This improved on previous algorithms for Planar Subgraph Isomorphism by Plehn and Voigt [20] of complexity $2^{O(k \log k)} \cdot n^{O(\sqrt{k})}$, and Alon et al. [2], of complexity $2^{O(k)} \cdot n^{O(\sqrt{k})}$ (which are not FPT algorithms). Finally, Dorn [8] improved the previous results and gave an algorithm of complexity $2^{O(k)} \cdot O(n)$. Eppstein [13, 15] also considered graphs of bounded genus, which generalize planar graphs. For an integer $g \geq 0$, a graph $G$ has genus at most $g$ if it can be embedded on an orientable surface of genus $g$,

---

*Humboldt University Berlin, Computer Science Department, Unter den Linden 6, 10099 Berlin, Germany. bonsma@informatik.hu-berlin.de. Date: June 14, 2018

i.e. a sphere with $g$ handles. In [15], an algorithm for Subgraph Isomorphism in bounded genus graphs of complexity $2^{O(k \log k)} \cdot O(n)$ is given. In addition, Eppstein [15] considered the even more general graph class of apex-minor free graphs, and gave an $f(k) \cdot O(n)$ time algorithm, where $f(k)$ is a rapidly growing function of $k$. We remark that for general graphs, FPT algorithms for Subgraph Isomorphism are unlikely [12]. The aforementioned results by Eppstein [13, 14, 15] in fact hold for the more general *counting version* of the problem, where the number of subgraphs of $G$ that are isomorphic to $P$ should be computed. In addition, in the case $P$ is connected, he gave an algorithm for *listing* all of these subgraphs in time $2^{O(k \log k)} \cdot O(n) + m \cdot k^{O(1)}$, where $m$ is the number of such subgraphs.

In this paper, we give an algorithm of complexity $2^{O(k)} \cdot O(n)$ for the counting version of Subgraph Isomorphism, for the case where $G$ has bounded genus. This generalizes the result for planar graphs in [8] and improves the complexity of the bounded genus result in [15]. In addition, we give an algorithm that lists all $m$ subgraphs isomorphic to $P$ in in time $2^{O(k)} \cdot O(n) + m \cdot k^{O(1)}$. This also holds for the case where $P$ is disconnected, and therefore answers an open question from Eppstein [13, 14]. This is achieved by using a simpler method for counting disconnected subgraphs. Our results hold for graphs of bounded non-orientable genus as well. For simplicity, we describe the orientable case only, and discuss the non-orientable case in Section 7. In Section 7 we also argue that our results apply to finding *induced* subgraphs.

There are many examples of problems that can be solved faster on planar graphs and other sparse graphs classes such as bounded genus graphs and $H$-minor free graphs, in particular in the area of FPT algorithms. For instance, for the aforementioned graph classes, many parameterized problems can be solved in *subexponential time* $2^{O(\sqrt{k})} \cdot O(n)$, see e.g. [1, 7, 4, 5]. The *theory of bidimensionality* [4, 5], easily gives subexponential time FPT algorithms for many problems restricted to the aforementioned graph classes. We note that this does however not apply to Subgraph Isomorphism. the reason is that bidimensionality applies to problems for which the existence of solutions does not change drastically when contracting edges. This is certainly not true for Subgraph Isomorphism, apart from a few very special cases for $P$, such as mainly (sets of) paths.

An essential ingredient for many of these algorithmic results on planar graphs, bounded genus graphs and $H$-minor free graphs is *dynamic programming over tree decompositions and branch decompositions*. (These are closely related; the minimum width of a tree decomposition and a branch decomposition differ by at most a small constant factor.) For problems such as finding a minimum independent set or maximum dominating set, straightforward algorithms of complexity $2^{O(w)} \cdot O(n)$ exist, where $w$ is the width of the given decomposition, see e.g. [19]. For more complex problems such as finding long paths and Subgraph Isomorphism, the best known dynamic programming algorithms have a complexity of $2^{O(w \log w)} \cdot O(n)$. However, when restricted to sparse graph classes, this can often be improved to $2^{O(w)} \cdot O(n)$. In the case of planar graphs, an essential tool is given by a special kind of branch decompositions, called *sphere cut decompositions*. These were introduced by Seymour and Thomas [22], and their algorithmic usefulness was first demonstrated by Dorn et al. [10, 11]. These have subsequently been applied many times for constructing fast algorithms. Loosely speaking, a branch decomposition for a graph $G$ consists of a labeled tree $T$, and every edge $e \in E(T)$ partitions the edges of $G$ into two graphs $G_1$ and $G_2$. In a sphere cut decomposition, for every $e \in E(T)$ a simple closed curve in the plane exists (a *noose*), that separates the plane into two regions, one containing $G_1$ and the other containing $G_2$. There are many problems that can be solved in time $2^{O(w \log w)} \cdot O(n)$ for general graphs on $n$ vertices, when a branch decomposition of width $w$ is given, but that can be solved in time $2^{O(w)} \cdot O(n)$ in the case of sphere cut decompositions [11]. It is generally believed that many algorithms for planar graphs can be extended to graphs of bounded genus. However, in the past, making this step has always been a complex technical task. For instance, Dorn et al. [9] consider the Hamilton Cycle problem and related problems on graphs of bounded genus, and reduce this case to the planar case by cutting the surface a number of times along nooses.

For the remaining planar case, dynamic programming over sphere cut decompositions is used, but this is relatively complex because the previous cuts need to be taken into account. Rué et al. [21] proposed a different dynamic programming method for graphs on surfaces: they define surface cut decompositions, where two subgraphs $G_1$ and $G_2$ defined by an edge of the branch decomposition are separated by a limited number of nooses, which have a limited number of common points.

**New techniques and overview of the paper** In Section 3 we give a dynamic programming algorithm for Subgraph Isomorphism that works for all graphs, when a branch decomposition is given. However, in order to give a good bound on its complexity, we restrict to bounded genus graphs and introduce a special kind of branch decomposition. *One of the main contributions of this paper is that we introduce surface split decompositions for graphs of bounded genus, and techniques for using these.* This is a type of branch decomposition that directly generalizes sphere cut decompositions. It allows algorithms that are significantly simpler than previous dynamic programming algorithms for bounded genus graphs. In fact, our algorithm and analysis is even significantly simpler than that of various previous algorithms for planar graphs (such as the previous algorithm for Planar Subgraph Isomorphism [8]). Informally, our basic but crucial observation is that for surfaces of higher genus, it is irrelevant that the two subgraphs $G_1$ and $G_2$ defined by an edge of the branch decomposition can share a complex boundary that consists of many nooses; it is only relevant that there are two disjoint (connected) regions $R_1$ and $R_2$ in the surface such that $G_i$ is drawn in $R_i$ for $i = 1, 2$. For a precise definition, and an example of how this property can be used for giving a good complexity bound, see Section 4. In Section 5 we give an algorithm for finding surface split decompositions in linear time, by generalizing a result by Tamaki [23] and Dorn [8], for finding low width sphere cut decompositions for planar graphs of bounded diameter. This is then applied in Section 6 to prove our main algorithmic results. We expect that the notion of surface split decompositions, our algorithm for finding them, and our technique for bounding the size of corresponding dynamic programming tables will be an impetus for the algorithmic research on bounded genus graphs. We believe that it will enable the generalization of various existing algorithmic results for planar graphs, and that it should allow for the simplification of various known results for bounded genus graphs, and more general graphs. In a subsequent paper, we will demonstrate this by applying our surface split decomposition techniques to other problems. We remark that algorithms that are conceptually simpler are not only convenient for the reader or programmer; often they are also faster in practice. Indeed, when restricted to planar graphs, we show in Section 7 that the hidden constants in the complexity bound of our algorithm are significantly smaller than those in the algorithm by Dorn [8] (although they are still huge). We start by giving definitions in Section 2.

## 2  Preliminaries

For basic graph theoretical notations not defined here we refer to [6]. The main graphs that we will consider throughout will be *simple*, but we will construct auxiliary graphs that may have parallel edges and loops, i.e. that may be *multi-graphs*. A *walk of length $k$* in a (multi-)graph $G$ is a sequence $v_0, e_1, v_1, e_2, \ldots, v_k$ where $v_i \in V(G)$, and $e_i \in E(G)$ is an edge of $G$ that is incident with $v_{i-1}$ and $v_i$. This is also called a walk *from $v_0$ to $v_k$*. Edges $e$ that are incident with $u$ and $v$ will also be denoted by $uv$ if there is no cause for confusion. We will also denote walks simply by their vertex sequence $v_0, v_1, \ldots, v_k$, which is unambiguous in simple graphs. A walk of length $k$ is a *cycle* or *$k$-cycle* if $v_0 = v_k$, the vertices $v_0, \ldots, v_{k-1}$ are distinct, and all edges are distinct. The *distance* from $u$ to $v$ is the minimum length over all walks from $u$ to $v$. The *eccentricity* of a vertex $u \in V(G)$ is the maximum distance from $u$ to $v$, over all $v \in V(G)$. By $d(v)$ we denote the *degree* of $v \in V(G)$, which is the number of incident edges.

The operation of *subdividing* an edge $e = uv$ consists of introducing a new vertex $x$ and edges $xu$ and $xv$, and deleting $e$. The reverse operation is called *suppressing the vertex $x$* (which can be done when $d(x) = 2$). An *isomorphism* between two simple graphs $G_1$ and $G_2$ is a bijective function $\phi : V(G_1) \to V(G_2)$ such that $uv \in E(G_1)$ if and only if $\phi(u)\phi(v) \in E(G_2)$. The *Subgraph Isomorphism (counting) problem* takes as input a simple (host) graph $G$ and a simple (pattern) graph $P$. The objective is to compute the number of subgraphs $G'$ of $G$ that are isomorphic to $P$. Such a subgraph $G'$ is called a *P-isomorph*, or *isomorph* if the graph $P$ is clear. Observe that the case where both $G$ and $P$ are non-simple is easily reduced to the simple case by subdividing every edge.

**Branch Decompositions** The set of leaves of a tree $T$ is denoted by $L(T)$. A *branch decomposition* of a graph $G$ is a tuple $(T, \mu)$ consisting of a ternary tree $T$ (i.e. all non-leaves have degree 3), and a bijection $\mu : L(T) \to E(G)$. For a subset $S \subseteq L(T)$, we will use $\mu(S)$ to denote the set of images. Every edge $e_T \in E(T)$ defines a *middle set* $mid(e_T) \subseteq V(G)$ in the following way: let $T_1 \subseteq V(T)$ and $T_2 \subseteq V(T)$ denote the vertex sets of the two tree components of $T - e_T$. The edge sets $\mu(L(T) \cap T_1)$ and $\mu(L(T) \cap T_2)$ partition the edges of $G$. By $G_1$ and $G_2$ we denote the respective subgraphs of $G$ induced by these edge sets. Now $mid(e_T)$ is defined as $V(G_1) \cap V(G_2)$. Hence this is the set of vertices that are both incident with edges belonging to $T_1$ and edges belonging to $T_2$. The *width* of $(T, \mu)$ is defined as

$$\text{width}(T, \mu) = \max_{e_T \in E(T)} |mid(e_T)|.$$

A *rooted* branch decomposition is a tuple $(T, \mu)$ where $T$ is a ternary tree, and a root $r \in L(T)$ is identified. In this case, $\mu$ is a bijection from $L(T) \backslash \{r\}$ to $E(G)$. A rooted branch decomposition can easily be obtained from a branch decomposition by subdividing an arbitrary edge with a new vertex $x$, and adding the vertex $r$ and edge $rx$. In the case of a rooted branch decomposition, for $e \in E(T)$, by $T_e \subseteq V(T)$ we denote the set of vertices of the component of $T - e$ that does not contain $r$. Similar to above, $T_e$ defines a subgraph of $G$, which is denoted by $G_e$. Since $T$ is ternary, every edge $e \in E(T)$ for which $T_e$ is non-trivial (i.e. not a single vertex) has two *children*; these are the other two edges of $T$ that are incident with the end vertex of $e$ in $T_e$. Observe that if $e_r \in E(T)$ is the edge incident with $r$, then $G_{e_r} = G$.

**Graphs embedded on surfaces** For an introduction to graphs embedded on surfaces we refer to [18]. Formally, a *surface* is a connected compact 2-manifold without boundary. For every integer $g \geq 0$, Let $\mathbb{S}_g$ denote a surface that is obtained by adding $g$ handles to a sphere. Hence $\mathbb{S}_g$ is an orientable surface of genus $g$, and in fact, for every orientable surface there is exactly one value of $g$ such that it is homeomorphic to $\mathbb{S}_g$ for [18, p. 82]. For ease of presentation, all surfaces that we consider will be orientable. In Section 7 we discuss how to extend our proofs to the non-orientable case. A *region* of a surface is a connected open set. The *boundary* of a region $R$ consists of all the points that lie in the closure of $R$ but not in $R$ itself. For a simple curve $C \subseteq \mathbb{S}_g$, all the points in $C$ that are not the end points are called *interior points*.

An *embedding* of a graph $G$ into the surface $\mathbb{S}_g$ consists of an injective mapping of the vertices $v$ of $G$ to points $\psi(v)$ in $\mathbb{S}_g$, and a mapping of edges $e = uv$ of $G$ to a simple curve $\psi(e)$ in $\mathbb{S}_g$ with end points $\psi(u)$ and $\psi(v)$, such that $\psi(e)$ shares no interior points with $\psi(e')$ or $\psi(v)$ for any $e' \in E(G) \backslash \{e\}$ or $v \in V(G)$. (That is, edges do not cross and only their end points overlap with vertices.) To simplify terminology and notation, if $\psi$ is an embedding of a graph $G$, then the images $\psi(v)$ and $\psi(e)$ (which are subsets of $\mathbb{S}_g$) for $v \in V(G)$ and $e \in E(G)$ will also be called *vertices* and *edges* of $G$, respectively. Let $X \subseteq \mathbb{S}_g$ be the union of all vertices and edges of an embedded graph $G$. The *faces* of $G$ are the maximal regions of $\mathbb{S}_g \backslash X$. An embedding is a *2-cell embedding* if every face is homeomorphic to an open disc. For a graph $G$ with $n$ vertices and $m$ edges, which is 2-cell embedded in $\mathbb{S}_g$ with $f$ faces, by *Euler's formula* it holds that $n - m + f = 2 - 2g$ [18, p. 85]. The boundary of every face $F$ of $G$ defines a closed walk $W$ in $G$ in a straightforward way, which will be called the *facial walk* for $F$. If the length of $W$ is $k$, then $F$ will be called a *$k$-face*.

Given a 2-cell embedding $\psi$ of a graph $G$ in $\mathbb{S}_g$, we can define the following two related (multi-) graphs, and their 2-cell embeddings in $\mathbb{S}_g$. The *dual graph* $G^*$ of $G$ has one vertex $v_F$ for every face $F$ of $G$, and one edge $e^*$ for every edge $e \in E(G)$, with end vertices $v_F$ and $v_{F'}$, where $F$ and $F'$ are the faces incident with $e$. Note that $G^*$ may have parallel edges and loops. $e^*$ is called the *dual edge* of $e$ and $e$ the *primal edge* of $e^*$. A straightforward 2-cell embedding $\psi^*$ of $G^*$ exists such that $v_F$ lies in the face $F$, and such that for $e \in E(G)$ and $e' \in E(G^*)$, $\psi(e)$ overlaps with $\psi^*(e')$ if and only if $e'$ is the dual edge of $e$.

The *radial graph* $\mathcal{R}_G$ of $G$ (which is also called the *vertex-face incidence graph*) is obtained by starting with the vertex set $V(G)$ (the *original vertices*), and adding a vertex $v_F$ for every face $F$ of $G$ (the *face vertices*). For every face $F$ in $G$ with corresponding facial walk $u_0, u_1, \ldots, u_{k-1}, u_0$, add $k$ edges $u_0 v_F, u_1 v_F, \ldots, u_{k-1} v_F$. Note that $\mathcal{R}_G$ may have parallel edges again. The original vertices and face vertices of $\mathcal{R}_G$ define a bipartition of $\mathcal{R}_G$. A straightforward 2-cell embedding $\psi^*$ of $\mathcal{R}_G$ exists such that for every face $F$ of $G$ it holds that $\psi^*(v_F) \in F$, for every $v \in V(G)$ it holds that $\psi^*(v) = \psi(v)$, and for every edge $e = uv_F$ of $\mathcal{R}_G$, interior points of $\psi^*(e)$ lie in the face $F$.

A *combinatorial embedding* $\pi$ of a graph $G$ consists of a cyclic order $\pi_v$ on the incident edges for every vertex $v \in V(G)$. From an embedding of a graph we obtain the corresponding combinatorial embedding by considering the clockwise order of edges around every vertex. A *map* is a connected graph $G$ together with a combinatorial embedding $\pi$. The set of all facial walks and therefore the number of faces of a map can easily be deduced (without constructing the embedding), so the number of faces $f$ of a map is well-defined. The *genus* $g$ of a map $G, \pi$ is the solution to $n - m + f = 2 - 2g$, where $n = |V(G)|$, $m = |E(G)|$ and $f$ is the number of faces of $G, \pi$. The *genus of a connected graph* $G$ is the minimum $g$ such that $G$ admits a combinatorial embedding of genus $g$. Given a map $G, \pi$ of genus $g$, a corresponding 2-cell embedding of $G$ in $\mathbb{S}_g$ exists. Our algorithms only work with combinatorial embeddings; it is not necessary to generate an embedding in $\mathbb{S}_g$. However for the correctness proofs it is convenient to argue with corresponding 2-cell embeddings of $G$ in $\mathbb{S}_g$.

## 3 A dynamic program for counting colorful isomorphs

In this section we give a dynamic programming algorithm for the following generalization of Subgraph Isomorphism. An instance of the *Colorful Subgraph Isomorphism* problem consists of a colored host graph $G$, and a pattern graph $P$. The *coloring* of $G$ is a function $\alpha : V(G) \to C$, with $C = \{1, \ldots, q\}$. This encodes a partition of $V(G)$ into $q$ sets. We remark that this is not required to be a *proper* vertex coloring, so adjacent vertices may receive the same color. A subgraph $G'$ of $G$ is called *colorful* if for every color $x \in C$, $G'$ contains a vertex $v$ of color $x$. (Note that $G'$ may have more than $q$ vertices.) The objective is to count the number of colorful $P$-isomorphs of $G$. We now present an algorithm for this problem. (When $q = 1$, this is the original counting problem.)

**Dynamic programming table** Let $(T, \mu)$ be a rooted branch decomposition of $G$. For every edge $e \in E(T)$, we will form a dynamic programming table $\mathcal{T}_e$. Informally, this table will store information about all possible subgraphs of the graph $G_e$, on at most $k$ vertices. Firstly, we distinguish between non-isomorphic subgraphs. Furthermore, subgraphs of $G_e$ that are isomorphic but intersect differently with the 'boundary' $\text{mid}(e)$ of $G_e$ are also considered distinct. Finally, we keep track of the set of colors that appear in these subgraphs. Two subgraphs of $G_e$ are only considered equivalent if they match in all three regards. In that case, there will be a single entry in the table that represents both subgraphs. We now define this formally. The definitions and dynamic programming update step are illustrated in Figure 1.

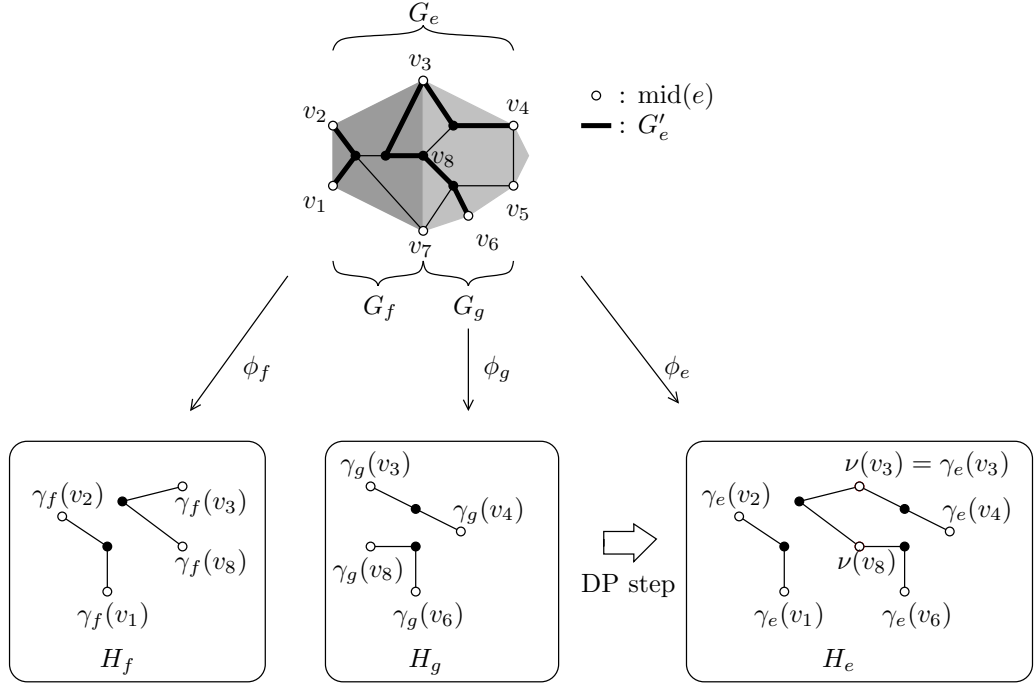

Figure 1: An illustration of the dynamic programming update step: An edge $e \in E(T)$ with children $f, g \in E(T)$ defines the subgraph $G_e$ of $G$. A subgraph $G'_e$ of $G_e$ is shown in bold. A table entry $(H_e, \gamma_e, A_e, \eta) \in \mathcal{T}_e$ that corresponds to $G'_e$ is obtained by combining the compatible entries $(H_f, \gamma_f, A_f, \eta_f) \in \mathcal{T}_f$ and $(H_g, \gamma_g, A_g, \eta_g) \in \mathcal{T}_g$. Vertex colors are ignored in the figure.

Let $H$ be a graph, and let $\gamma$ be a mapping from $\mathrm{mid}(e)$ to $V(H) \cup \{\mathrm{nil}\}$, which is *injective on* $V(H)$. To be precise, every vertex of $V(H)$ occurs at most once as a $\gamma$-image, but multiple vertices may be mapped to nil. Furthermore, let $A \subseteq \{1, \ldots, q\}$. For such a tuple $(H, \gamma, A)$, a subgraph $G'$ of $G_e$ is called an $(H, \gamma, A)$-*subgraph* if the following two properties hold.

- There is an isomorphism $\phi : V(G') \to V(H)$ with $\gamma(v) = \phi(v)$ for all $v \in \mathrm{mid}(e) \cap V(G')$, and $\gamma(v) = \mathrm{nil}$ for all $v \in \mathrm{mid}(e) \setminus V(G')$.

- For all colors $x \in C$, $x \in A$ if and only if $G'$ contains a vertex of color $x$.

For $e \in E(T)$, the *dynamic programming table* $\mathcal{T}_e$ will now contain *entries* $(H, \gamma, A, \eta)$, where $H$, $\gamma$ and $A$ are as defined above, and $\eta$ is a non-negative integer. The idea is that such a table entry indicates that $G_e$ contains exactly $\eta$ non-equivalent $(H, \gamma, A)$-subgraphs. Two table entries $(H_1, \gamma_1, A_1, \eta_1)$ and $(H_2, \gamma_2, A_2, \eta_2)$ are *equivalent* if the following properties hold.

- There is an isomorphism $\phi : V(H_1) \to V(H_2)$ such that for all $v \in \mathrm{mid}(e)$, either $\gamma_1(v) = \gamma_2(v) = \mathrm{nil}$, or $\phi(\gamma_1(v)) = \gamma_2(v)$ holds.

- $A_1 = A_2$.

Observe that the above definition satisfies the following property: a subgraph $G'$ of $G_e$ is both a $(H_1, \gamma_1, A_1)$-subgraph and a $(H_2, \gamma_2, A_2)$-subgraph if and only if $(H_1, \gamma_1, A_1, \eta_1)$ and $(H_2, \gamma_2, A_2, \eta_2)$ are equivalent. Therefore, when two entries $(H_1, \gamma_1, A_1, \eta_1)$ and $(H_2, \gamma_2, A_2, \eta_2)$ are equivalent, we can *merge* them by replacing them with the single entry $(H_1, \gamma_1, A_1, \eta_1 + \eta_2)$. We say that the table $\mathcal{T}_e$ is $k$-*correct* if

- for every tuple $(H, \gamma, A)$, $\mathcal{T}_e$ contains an entry $(H, \gamma, A, \eta)$ if and only if $G_e$ contains exactly $\eta \geq 1$ graphs $G'$ with $|V(G')| \leq k$ that are $(H, \gamma, A)$-subgraphs, and

- $\mathcal{T}_e$ contains no pairs of equivalent entries.

**Dynamic programming update step**  Let $e \in E(T)$ be an edge with children $f$ and $g$. We will now show how a $k$-correct table $\mathcal{T}_e$ for $e$ can be obtained from $k$-correct tables $\mathcal{T}_f$ and $\mathcal{T}_g$ for $f$ and $g$, respectively. Let $G'_f$ and $G'_g$ be subgraphs of $G_f$ and $G_g$, respectively. They are called *compatible* if for every $v \in \mathrm{mid}(f) \cap \mathrm{mid}(g)$ it holds that $v \in V(G'_f)$ if and only if $v \in V(G'_g)$. Note that the function $\psi(G'_f, G'_g) := \left(V(G'_f) \cup V(G'_g), E(G'_f) \cup E(G'_g)\right)$ is a *bijection* between pairs of compatible subgraphs of $G_f$ and $G_g$ and subgraphs of $G_e$. Similarly, we define two entries $(H_f, \gamma_f, A_f, \eta_f) \in \mathcal{T}_f$ and $(H_g, \gamma_g, A_g, \eta_g) \in \mathcal{T}_g$ to be *compatible* if for all $v \in \mathrm{mid}(f) \cap \mathrm{mid}(g)$, it holds that $\gamma_f(v) = \mathrm{nil}$ if and only if $\gamma_g(v) = \mathrm{nil}$. We now show how to *combine* two such compatible entries into an entry $(H_e, \gamma_e, A_e, \eta_e)$ that satisfies the following property: if $G'_f$ is an $(H_f, \gamma_f, A_f)$-subgraph and $G'_g$ is an $(H_g, \gamma_g, A_g)$-subgraph, then $\psi(G'_f, G'_g)$ is an $(H_e, \gamma_e, A_e)$-subgraph.

- For all $v \in \mathrm{mid}(f) \cap \mathrm{mid}(g)$ with $\gamma_f(v) \neq \mathrm{nil}$ (and thus $\gamma_g(v) \neq \mathrm{nil}$): identify the vertex $\gamma_f(v)$ of $H_f$ with the vertex $\gamma_g(v)$ of $H_g$, and call the new vertex $\nu(v)$. This gives the graph $H_e$.

- For all $v \in \mathrm{mid}(e)$:
  1. If $v \in \mathrm{mid}(f) \setminus \mathrm{mid}(g)$: set $\gamma_e(v) = \gamma_f(v)$.
  2. If $v \in \mathrm{mid}(g) \setminus \mathrm{mid}(f)$: set $\gamma_e(v) = \gamma_g(v)$.
  3. If $v \in \mathrm{mid}(g) \cap \mathrm{mid}(f)$: set $\gamma_e(v) = \nu(v)$.

  By definition of $\mathrm{mid}(e)$, this covers all cases and thus defines the function $\gamma_e$.

- Set $A_e = A_f \cup A_g$.

- Set $\eta_e := \eta_f \cdot \eta_g$.

The above definitions and observations show that if there are $\eta_f$ $(H_f, \gamma_f, A_f)$-subgraphs in $G_f$, and $\eta_g$ $(H_g, \gamma_g, A_g)$-subgraphs in $G_g$, then there are $\eta_f \cdot \eta_g$ $(H_e, \gamma_e, A_e)$-subgraphs in $G_e$ that are the result of combining graphs of the former two types in every possible combination. However, there may also be $(H_e, \gamma_e, A_e)$-subgraphs of $G_e$ that are the result of combining different types of subgraphs of $G_f$ and $G_g$. Therefore, the entire dynamic programming update step requires the merging of entries as well. We will now present this in detail.

Assuming that we have $k$-correct tables $\mathcal{T}_f$ and $\mathcal{T}_g$ for $f$ and $g$ respectively, we construct a $k$-correct table $\mathcal{T}_e$ for $e$ as follows: We start with an empty table $\mathcal{T}_e$. Then we consider every pair of compatible entries from $\mathcal{T}_f$ and $\mathcal{T}_g$, and combine them as described above. For every combination, this yields a possible entry $(H, \gamma, A, \eta)$ for $\mathcal{T}_e$. In case that $H$ has more than $k$ vertices, we ignore this possible entry. Otherwise, we check whether $\mathcal{T}_e$ already contains an equivalent entry $(H', \gamma', A', \eta')$. If so, we merge the two entries. If not, we add the entry $(T, \gamma, A, \eta)$ to the table $\mathcal{T}_e$. Then we continue with the next pair of compatible entries from $\mathcal{T}_f$ and $\mathcal{T}_g$.

**Lemma 1** *Let $(T, \mu)$ be a rooted branch decomposition for a colored graph $G$, and $k$ be an integer. Let $e \in E(T)$ be an edge with children $f$ and $g$, for which $k$-correct tables $\mathcal{T}_f$ and $\mathcal{T}_g$ are given. Then the table $\mathcal{T}_e$ that is constructed with the above dynamic programming update step is $k$-correct for $e$. The construction takes time $X^3 \cdot f(k) \cdot k^{O(1)}$, where $f(k)$ is the complexity of deciding whether two entries are equivalent, and $X$ is an upper bound on the size of a $k$-correct table.*

*Proof:* We argue that the construction of $\mathcal{T}_e$ is correct. Clearly, $\mathcal{T}_e$ contains no pair of equivalent entries. First we show that for every resulting entry $(H_e, \gamma_e, A_e, \eta_e) \in \mathcal{T}_e$, there are at least $\eta_e$ $(H_e, \gamma_e, A_e)$-subgraphs in $G_e$. Since $\mathcal{T}_f$ is $k$-correct, every entry $(H_f, \gamma_f, A_f, \eta_f) \in \mathcal{T}_f$ corresponds to a set of $\eta_f$ $(H_f, \gamma_f, A_f)$-subgraphs of $G_f$. A similar statement holds for $\mathcal{T}_g$. Consider compatible entries $(H_f, \gamma_f, A_f, \eta_f) \in \mathcal{T}_f$ and $(H_g, \gamma_g, A_g, \eta_g) \in \mathcal{T}_g$. The bijection $\psi$ and the construction of $(H_e, \gamma_e, A_e, \eta_e)$ show that there are $\eta_f \cdot \eta_g$ subgraphs of $G_e$ that are $(H_e, \gamma_e, A_e)$-subgraphs, that are obtained by combining every $(H_f, \gamma_f, A_f)$-subgraph with every $(H_g, \gamma_g, A_g)$-subgraph. When merging a new entry $(H_e, \gamma_e, A_e, \eta_e)$ with an existing entry $(H'_e, \gamma'_e, A'_e, \eta'_e)$, the corresponding two sets of subgraphs of $G_e$ are disjoint, because they correspond to different types of subgraphs of $G_f$ and $G_g$. This shows that for every entry $(H_e, \gamma_e, A_e, \eta_e) \in \mathcal{T}_e$, $\eta_e$ is a lower bound for the number of $(H_e, \gamma_e, A_e)$-subgraphs in $G_e$.

We conclude the correctness proof by showing that every graph is counted this way. Consider a subgraph $G'_e$ of $G_e$ on at most $k$ vertices. Its restrictions to $G_f$ and $G_g$ respectively give a compatible pair $G'_f$ and $G'_g$, both with at most $k$ vertices. Since $\mathcal{T}_f$ and $\mathcal{T}_g$ $k$-correct, there are unique entries $(H_f, \gamma_f, A_f, \eta_f) \in \mathcal{T}_f$ and $(H_g, \gamma_g, A_g, \eta_g) \in \mathcal{T}_g$ such that $G'_f$ is one of the $\eta_f$ $(H_f, \gamma_f, A_f)$-subgraphs of $G_f$, and a similar statement holds for $G'_g$. This shows that the subgraph $G'_e$ of $G_e$ is counted by the dynamic programming update step; it is one of the $\eta_f \cdot \eta_g$ resulting subgraphs.

Finally, we consider the complexity. Both $\mathcal{T}_f$ and $\mathcal{T}_g$ contain at most $X$ entries, and the same holds for $\mathcal{T}_e$, at every time during its construction. Therefore, there are at most $X^2$ pairs of entries from $\mathcal{T}_f$ and $\mathcal{T}_g$ that should be considered. Testing whether they are compatible, and if so, combining them can be done in polynomial time $k^{O(1)}$. Subsequently, there are at most $X$ entries of $\mathcal{T}_e$ that should be tested for equivalence, which takes times $f(k) \cdot X$. This proves the stated complexity[1]. □

This dynamic programming update step is the core of the following dynamic programming algorithm:

- First, for every edge $e \in E(T)$ that has no children, $G_e$ consists of a single edge, so it is trivial to compute a $k$-correct table $\mathcal{T}_e$. To be precise, assuming that $\mathrm{mid}(e) \neq \emptyset$, the five entries of $\mathcal{T}_e$ correspond to the empty graph, a graph consisting of a single edge, two graphs consisting of a single vertex (with different $\gamma$), and one graph on two vertices, without edges.

- For every edge $e \in E(T)$ with two children $f$ and $g$, we compute a $k$-correct table $\mathcal{T}_e$ using the dynamic programming update step, after $k$-correct tables for $\mathcal{T}_f$ and $\mathcal{T}_g$ have been computed.

- After computing $k$-correct tables for all edges of $T$, we inspect the table $\mathcal{T}_{e_r}$ where $e_r$ is the root edge of $(T, \mu)$. Since $G_{e_r} = G$, $\mathrm{mid}(e_r) = \emptyset$. Therefore, $\mathcal{T}_{e_r}$ contains at most one entry $(H, \gamma, A, \eta)$ such that $H$ is isomorphic to $P$ and $A = \{1, \ldots, q\}$. If there is such an entry $(H, \gamma, A, \eta)$, we return $\eta$, and otherwise we return 0.

The correctness of the above procedure follows from the definitions. Combined with Lemma 1 this gives the following theorem. Note that a branch decomposition $(T, \mu)$ of a graph on $m$ edges has $|E(T)| \in O(m)$.

**Theorem 2** *Let $(T, \mu)$ be a rooted branch decomposition of a colored graph $G$ with $m$ edges. In time $X^3 \cdot f(k) \cdot k^{O(1)} \cdot O(m)$, it can be computed how many colorful subgraphs of $G$ are isomorphic to a given graph $P$ on $k$ vertices, where $f(k)$ is the complexity of deciding whether two entries are equivalent, and $X$ is an upper bound on the size of a $k$-correct table.*

The above theorem applies to general graphs, for which an appropriate bound for $X$ can be given. However, to obtain the desired complexity, we introduce surface split decompositions in the next section. In the case that an embedding $\pi$ of $G$ of bounded genus is given, and $(T, \mu)$ is a surface split decomposition, we give a good upper bound for $X$.

---

[1] We remark that there is room for improvements in the the complexity (constant factors) here: when the tables are ordered, not all entry combinations have to be tested for compatibility, and not all entries have to be considered for testing equivalence. See e.g. [19] for examples of such improvements. For simplicity we ignore this in our analysis.

## 4 Surface split decompositions and a bound for the table size

In this section, we define surface split decompositions for graphs embedded in $\mathbb{S}_g$. These generalize the notion of sphere cut decompositions, which are defined for graphs embedded on the sphere $\mathbb{S}_0$ (i.e. plane graphs). Given an embedding of $G$ in the sphere $\mathbb{S}_0$, a *noose* is a simple closed curve $N$ in $\mathbb{S}_0$ that intersects $G$ only in its vertices. Hence by the Jordan Curve Theorem, $\mathbb{S}_0 \setminus N$ consists of exactly two connected components, the regions $R_1$ and $R_2$. A branch decomposition $(T, \mu)$ of $G$ is a *sphere cut decomposition* [11] if for every $e_T \in E(T)$ with corresponding subgraphs $G_1$ and $G_2$ of $G$, there exists a noose $N$ that divides $\mathbb{S}_0$ into $R_1$ and $R_2$, such that for $i = 1, 2$, $G_i$ lies in $N \cup R_i$, and $N$ intersects $G$ exactly in mid$(e)$. Our simple but essential observation is that the fact that the two regions $R_1$ and $R_2$ are separated by a simple closed curve $N$ is actually irrelevant; this is merely an artefact of restricting to the sphere. We will show that the appropriate way of generalizing sphere cut decompositions to surfaces of higher genus is surprisingly simple, as follows.

**Definition 3** *Let $G$ be a graph embedded on $\mathbb{S}_g$. A branch decomposition $(T, \mu)$ of $G$ is called a* surface split decomposition *if for every $e \in E(T)$ and corresponding subgraphs $G_1$ and $G_2$ of $G$, there are disjoint regions $R_1 \subseteq \mathbb{S}_g$ and $R_2 \subseteq \mathbb{S}_g$ such that for $i = 1, 2$, all vertices and edges of $G_i$ are drawn entirely in the closure of $R_i$.*

Observe that this definition implies that all vertices in mid$(e)$ lie on the boundary of the closure of $R_1$ and on the boundary of the closure of $R_2$. Note also that w.l.o.g. we may assume that all other vertices of $G_i$ and internal points of edges of $G_i$ lie entirely in $R_i$ itself (not just in the closure). We stress that it is crucial that $R_1$ and $R_2$ are connected *open* sets. If not, then firstly the above definition is not a generalization of sphere cut decompositions, but more importantly, the proof of Lemma 7 below fails. It is essential that there is no single point $x \in R_i$ such that $R_i \setminus \{x\}$ consists of multiple connected components (for $i = 1, 2$). Finally, we remark that even if the regions $R_1$ and $R_2$ are chosen such that their boundaries are the same, this boundary is not necessarily a simple curve if $g \geq 1$. This is illustrated in the figures below. The definition easily extends to *maps* $G, \pi$ of genus $g$: $(T, \mu)$ is a *surface split decomposition* of $G, \pi$ if it is a surface split decomposition for any embedding of $G$ in $\mathbb{S}_g$ that corresponds to the combinatorial embedding $\pi$.

Let $(T, \mu)$ be a surface split decomposition for a map $G, \pi$. In this section, we firstly give a bound on the size of a $k$-correct table, using the fact that $(T, \mu)$ is a surface split decomposition. Secondly, we show that equivalent pairs can be recognized in polynomial time.

For the bound on the table size, we will use a bound on the number of different graphs on $n$ vertices, embedded in a surface of genus $g$. To be precise, we will consider *simple, connected* graphs $G$, that come with a combinatorial embedding $\pi$. Such a pair $G, \pi$ is called a *simple map*. In addition, a tuple $(u, v)$ of vertices is given such that $uv \in E(G)$. This is the *root*, in Figure 2 indicated with an arc. Such a combination $G, \pi, (u, v)$ is called a *simple rooted map*. Two simple rooted maps $G, \pi, (u, v)$ and $G', \pi', (u', v')$ are *equivalent* if there is an isomorphism $\phi : V(G) \to V(G')$ with $\phi(u) = u'$, $\phi(v) = v'$, and that preserves the clockwise order of edges around every vertex, or reverses all of these orders. In other words, $\phi$ should map faces (facial walks) of $G$ to faces of $G'$. In case an edge labeling is given for both simple rooted maps, we also require that $\phi$ maps edges to edges of the same label. To bound the number of simple rooted maps, we apply a result by Bender and Canfield [3][2].

---

[2]Bender and Canfield included multi-graphs (i.e. *rooted maps*) in their bound, but clearly this yields an upper bound for the number of *simple* rooted maps.

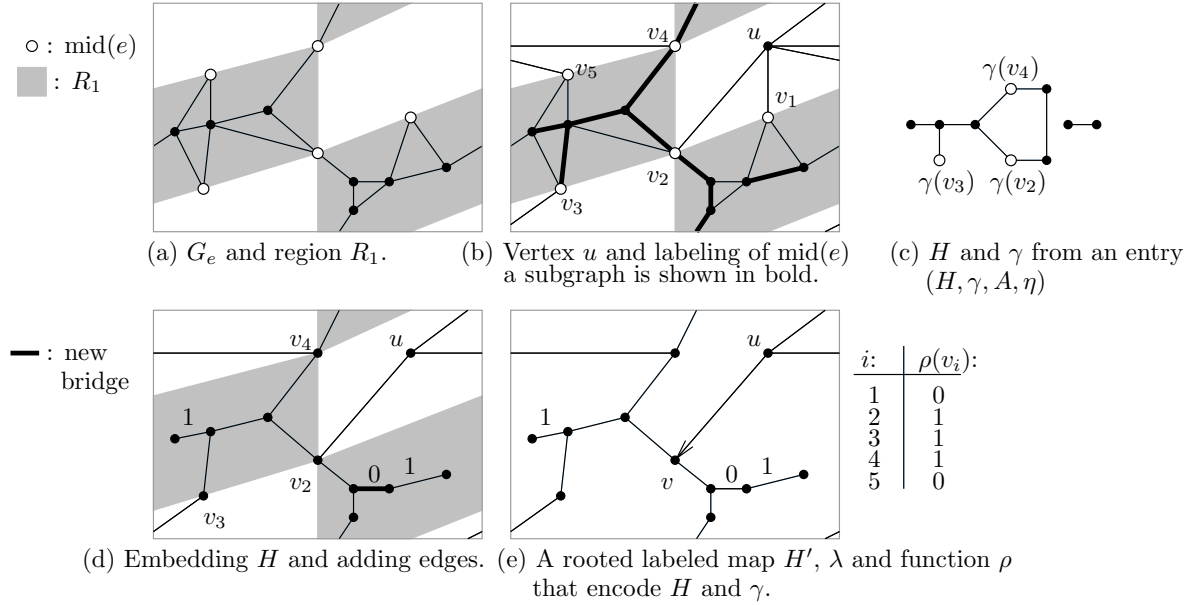

(a) $G_e$ and region $R_1$.

(b) Vertex $u$ and labeling of mid($e$) a subgraph is shown in bold.

(c) $H$ and $\gamma$ from an entry $(H, \gamma, A, \eta)$

(d) Embedding $H$ and adding edges.

(e) A rooted labeled map $H'$, $\lambda$ and function $\rho$ that encode $H$ and $\gamma$.

Figure 2: An illustration of the proof of Lemma 7. The rectangle represents a torus; edges that leave the rectangle on the left continue at the same height on the right, and a similar statement holds for edges leaving the top and bottom of the rectangle.

**Theorem 4 ([3])** *There are $O(m^{5(g-1)/2} 12^m)$ equivalence classes of simple rooted maps of genus $g$ on $m$ edges.*

Using an elementary argument based on Euler's formula, one can prove that the following bound holds in the case of simple graphs.

**Proposition 5** *Let $G$ be a simple graph of genus $g$, with $n$ vertices and $m$ edges. Then $m \leq 3n - 6 + 6g$.*

Combining these bounds yields the following rough upper bound in terms of the number of vertices.

**Corollary 6** *Let $g$ be a constant. There are $2^{O(n)}$ simple rooted maps of genus at most $g$ on $n$ vertices.*

*Proof:* When considering $g$ to be constant, Theorem 4 and Proposition 5 show that the number of simple rooted maps on $n$ vertices of genus $g$ can be bounded by

$$O\left((3n + 6g)^{5(g-1)/2} \cdot 12^{3n} \cdot 12^{6g}\right) \subseteq n^{O(g)} \cdot 2^{O(n)} \subseteq 2^{O(n)}.$$

Clearly, this is then also an upper bound for the number of simple rooted maps of genus *at most* $g$. □

**Lemma 7** *Let $g$ be a constant. Let $G, \pi$ be a simple map of genus at most $g$, for which a coloring with $q$ colors and a surface split decomposition $(T, \mu)$ of width $w$ are given. Let $k$ be an integer. For $e \in E(T)$, let $\mathcal{T}_e$ be a $k$-correct table. Then $|\mathcal{T}_e| \in 2^q \cdot 2^w \cdot 2^{O(k)}$.*

*Proof:* For all entries $(H, \gamma, A, \eta) \in \mathcal{T}_e$ where $H$ has at most $k$ vertices, we will *encode* $H$ and $\gamma$ by a simple rooted map $H', \pi', (u, v)$ on at most $k+1$ vertices, together with a 0/1-labeling $\lambda$ of a subset of the edges of $H'$, and a 0/1-labeling $\rho$ of the vertices in mid($e$). This means that any two non-equivalent entries $(H_1, \gamma_1, A_1, \eta_1)$ and $(H_2, \gamma_2, A_2, \eta_2)$ either have $A_1 \neq A_2$, or yield a different labeling $\rho$, or yield non-equivalent rooted maps. The encoding is illustrated in Figure 2.

We use the following auxiliary graph. Since $(T, \mu)$ is a surface split decomposition, there are disjoint regions $R_1$ and $R_2$ in $\mathbb{S}_g$, such that $G_e$ lies in the closure of $R_1$, and mid$(e)$ lies on the boundary of both $R_1$ and $R_2$. Thus we can extend $G_e$ by drawing a vertex $u$ in $R_2$, and drawing edges in the closure of $R_2$ from $u$ to every vertex in mid$(e)$, while maintaining an embedding in $\mathbb{S}_g$. Number the vertices of mid$(e)$ $v_1, \ldots, v_t$, corresponding to the clockwise order of edges around $u$ [3].

Now we show how to construct the encoding $H', \pi', \rho, \lambda$ for an entry $(H, \gamma, A, \eta) \in \mathcal{T}_e$. Firstly, for all vertices $v \in$ mid$(e)$, set $\rho(v) = 0$ if and only if $\gamma(v) = $ nil. The graph $H'$ is constructed as follows. Since $\mathcal{T}_e$ is $k$-correct, $H$ corresponds to a subgraph of $G_e$, so $H$ can also be drawn in the closure of $R_1$, such that all vertices that are $\gamma$-images are drawn on the boundary of $R_1$. Start with such an embedding. Next, add the vertex $u$, and edges $uv_i$ for every $v_i \in$ mid$(e)$ with $\gamma(v_i) \neq $ nil. Draw these as described in the previous paragraph. This yields a simple graph embedded in $\mathbb{S}_g$, on at most $k+1$ vertices. However, it may not be connected (if $H$ has components that contain no $\gamma$-image). Add edges between different components until the graph is connected. This yields $H'$. Clearly, drawing these new edges can be done while maintaining an embedding. Hence $H'$ is embedded in $\mathbb{S}_g$, and the corresponding combinatorial embedding $\pi'$ has genus at most $g$. To obtain a *rooted map*, choose $i$ to be the lowest index such that $\rho(v_i) = 1$. Then the tuple $(u, v_i)$ is chosen as the root of $H', \pi'$. That is, $v = v_i$. A *bridge* of a connected graph $G$ is an edge $e \in E(G)$ such that $G - e$ is disconnected. For all bridges $e \in E(H')$, we set $\lambda(e) = 1$ if $e \in E(H)$, and $\lambda(e) = 0$ otherwise.

To argue that this is an encoding, we show that the pair $H, \gamma$ can be reconstructed from $H', \pi', (u, v), \rho, \lambda$, without using information about vertex labels of $H'$ (since we apply a bound on equivalence classes of maps, or in other words, unlabeled maps). Let $i$ be the lowest index such that $\phi(v_i) = 1$, and let $(u, v)$ be the root of $H'$. Then $\gamma(v_i) = v$. Let $i'$ be the next index such that $\phi(v_{i'}) = 1$, and let $(u, v')$ the the next edge incident with $u$, in clockwise order. Then $\gamma(v_{i'}) = v'$. Continuing this way, and setting $\gamma(v_j) = $ nil for all $v_j$ with $\rho(v_j) = 0$, we obtain $\gamma$. Next, delete $u$ and every bridge $e$ with $\lambda(e) = 0$. This way we obtain the graph $H$. This yields the following conclusion: Suppose two entries $(H_1, \gamma_1, A_1, \eta_1)$ and $(H_2, \gamma_2, A_2, \eta_2)$ yield encodings $H'_1, \pi'_1, (u_1, v_1), \rho_1, \lambda_1$ and $H'_2, \pi'_2, (u_2, v_2), \rho_2, \lambda_2$ respectively, such that $\rho_1 = \rho_2$ and such that the edge labeled simple rooted maps $H'_1, \pi'_1, (u_1, v_1)$ and $H'_2, \pi'_2, (u_2, v_2)$ are equivalent. (Recall that this requires that the isomorphism maps edges of to edges with the same label.) Suppose that in addition $A_1 = A_2$. Then $(H_1, \gamma_1, A_1, \eta_1)$ and $(H_2, \gamma_2, A_2, \eta_2)$ are equivalent.

Since a $k$-correct table contains no pairs of equivalent entries, we can now give an upper bound for its size. There are at most $2^w$ possibilities for $\rho$, at most $2^q$ possibilities for $A$, at most $2^{O(k)}$ simple rooted maps on at most $k+1$ vertices of genus at most $g$ (Corollary 6), and at most $2^k$ possible labelings $\lambda$ (since a graph on $k+1$ vertices contains at most $k$ bridges.) Therefore, the number of entries in a $k$-correct table is bounded by $2^q \cdot 2^w \cdot 2^{O(k)}$. □

Now we show how to test whether two table entries are equivalent. This is based on the algorithm by Miller [17], for testing whether two graphs of bounded genus are isomorphic. See [16] for an alternative algorithm with the same complexity.

**Theorem 8 ([17])** *In time $n^{O(g)}$, it can be tested whether two graphs on $n$ vertices of genus $g$ are isomorphic.*

---

[3] We remark that if $R_2$ is not homeomorphic to a disc, then there are multiple ways of drawing these edges, resulting in different clockwise orders. Some will yield a better bound than others, but for proving the stated upper bound that is not important; it is only important that we choose one way of drawing these edges and use this for every entry of $\mathcal{T}_e$. We remark also that at this point we require that $R_2$ is not only connected, but also *open*, in order to draw the edges.

**Proposition 9** *Let $(T, \mu)$ be a branch decomposition of a map $G, \pi$ of genus $g$, and let $\mathcal{T}_e$ be a $k$-correct dynamic programming table for $e \in E(T)$. In time $k^{O(g)}$, it can be decided whether two entries $(H_1, \gamma_1, A_1, \eta_1)$ and $(H_2, \gamma_2, A_2, \eta_2)$ of $\mathcal{T}_e$ are equivalent.*

*Proof:* We assume that

- $H_1$ and $H_2$ have the same number of vertices,
- for every $v \in \text{mid}(e)$, $\gamma_1(v) \in V(H_1)$ if and only if $\gamma_2(v) \in V(H_2)$,
- for every $v \in \text{mid}(e)$ with $\gamma_1(v) \neq \text{nil}$, $\gamma_1(v)$ and $\gamma_2(v)$ have the same degree, and
- $A_1 = A_2$.

Otherwise, we may immediately conclude that the entries are not equivalent. We shall make small modifications to both $H_1$ and $H_2$, to ensure that any isomorphism $\phi$ of the resulting graphs $H_1'$ and $H_2'$ maps $\gamma_1(v)$ to $\gamma_2(v)$ for every $v \in \text{mid}(e)$. Then it suffices to test whether $H_1'$ and $H_2'$ are isomorphic.

Number the vertices $v \in \text{mid}(e)$ with $\gamma_1(v) \neq \text{nil}$ as $v_1, \ldots, v_t$, in an arbitrary order. We modify $H_1$ by adding $k$ vertices, all joined by an edge to $\gamma_1(v_1)$, and adding $2k$ vertices joined to $\gamma_1(v_2)$, etc. This ensures that in the resulting graph $H_1'$, for all $i$, the vertex $\gamma_1(v_i)$ has a unique degree. Apply the same modification to $H_2$ to obtain $H_2'$. Since the table $\mathcal{T}_e$ is $k$-correct, both $H_1$ and $H_2$ are isomorphic to subgraphs of $G$, and therefore have genus at most $g$. Adding degree 1 vertices does not increase the genus, so $H_1'$ and $H_2'$ have genus at most $g$ as well.

Because of the unique degrees, if there exists an isomorphism $\phi : V(H_1') \to V(H_2')$, then for all $i$, $\phi(\gamma_1(v_i)) = \gamma_2(v_i)$. It follows that $H_1'$ and $H_2'$ are isomorphic if and only if $(H_1, \gamma_1)$ and $(H_2, \gamma_2)$ are equivalent. The former can be decided in time $x^{O(g)}$ (Theorem 8), where $x = |V(H_1')|$. Since $x \in O(k^3)$, this shows that we can test equivalence in time $k^{3 \cdot O(g)} = k^{O(g)}$. □

Combining Theorem 2 with Lemma 7 and Proposition 9 gives the following theorem. Here we use that a graph on $n$ vertices of genus $g$ has at most $O(n+g)$ edges, and that $k^{O(1)} \cdot 2^{O(k)} \subseteq 2^{O(k)}$.

**Theorem 10** *Let $g$ be a constant. Let $G, \pi$ be a simple map on $n$ vertices of genus at most $g$, for which a coloring with $q$ colors and a surface split decomposition $(T, \mu)$ of width $w$ are given. Let $P$ be a graph on $k$ vertices. In time $8^q \cdot 8^w \cdot 2^{O(k)} \cdot O(n)$, it can be computed how many colorful subgraphs of $G$ are isomorphic to $P$.*

## 5 Constructing surface split decompositions

Tamaki [23] gave a linear time algorithm for constructing a branch decomposition of width $2d+1$ of a graph $G$ embedded in the sphere, when a vertex $r \in V(G)$ of eccentricity $d$ is given[4]. Dorn [8] gave a different presentation of the construction and showed that it yields in fact a sphere cut decomposition. We show how the construction of [23, 8] can be extended to surfaces of arbitrary genus, and prove the following theorem.

**Theorem 11** *Let $G, \pi$ be a map of genus $g$, for which a vertex $r \in V(G)$ of eccentricity $d$ is given. In linear time, we can construct a surface split decomposition $(T, \mu)$ of $G$ of width at most $(2g+1)(4d+3)/2$.*

Note that for $g = 0$, Theorem 11 implies the aforementioned result from [23] and [8] since $\lfloor (4d+3)/2 \rfloor = 2d+1$. The remainder of this section will be devoted to the proof of Theorem 11. Before we construct the surface split decomposition $(T, \mu)$, we will construct a number of auxiliary graphs. These are illustrated in Figure 3.

---
[4]Tamaki formulates the result differently, such that $r$ corresponds to the outer face of a graph embedded in the plane.

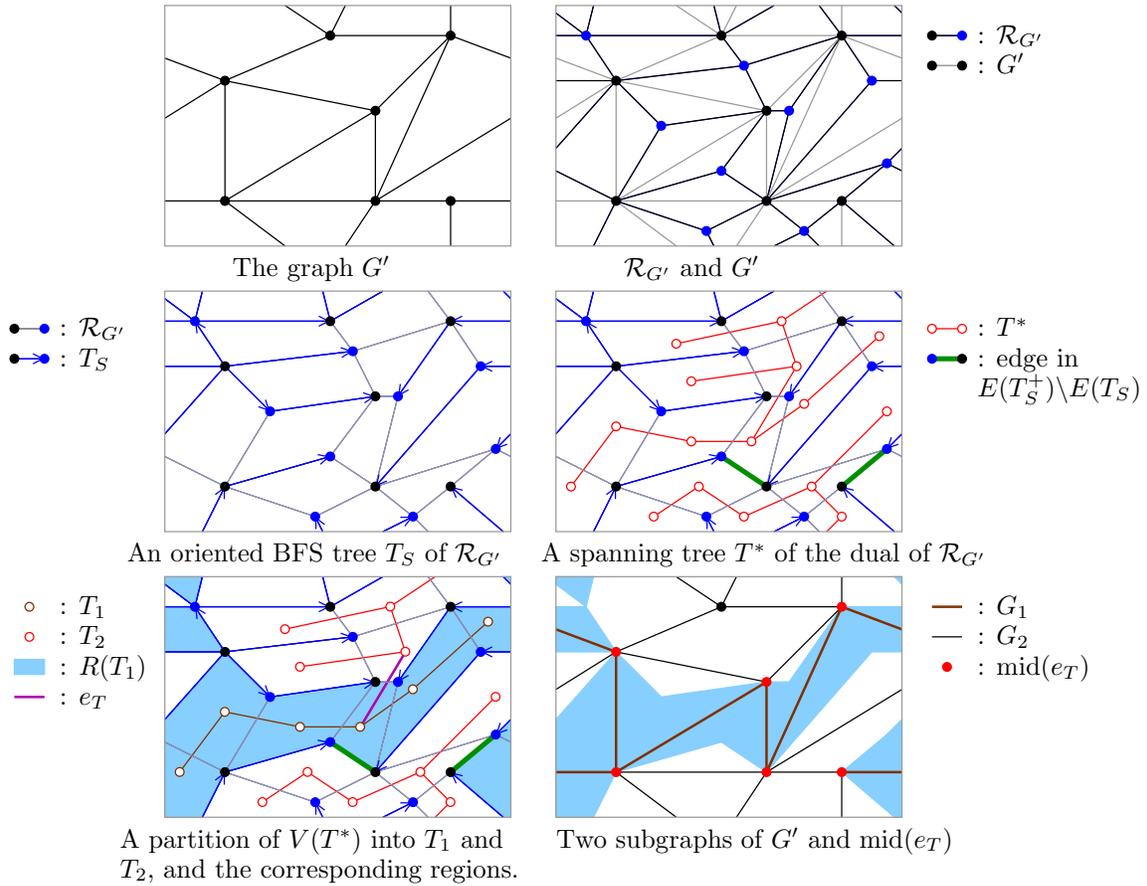

Figure 3: An illustration of the graphs used for the proof of Theorem 11, for the case of $g = 1$. The rectangle represents a torus; edges that leave the rectangle on the left continue at the same height on the right, and a similar statement holds for edges leaving the top and bottom of the rectangle.

The objective of the *first stage* of the construction is to construct a tree $T^*$, such that for every $e \in E(G)$, there is a unique vertex of $T^*$ associated with $e$. This can be thought of as the function $\mu$, from a subset of $V(T^*)$ to $E(G)$. We will show in Lemmas 12 and 14 below that this $T^*$ is already 'almost' the desired surface split decomposition. However, $T^*$ will not yet be ternary (though it will have maximum degree 3), and the vertices that are associated with edges of $G$ are not necessarily leaves. This is subsequently addressed in the *second stage* of the construction.

**First stage of the construction** First we modify the graph $G$ as follows: for every edge $e = uv$, add two extra parallel edges between $u$ and $v$, one on either side of $e$ (while maintaining a combinatorial embedding). This ensures that all original edges of $G$ are incident with two 2-faces, and that every vertex has degree at least 3. Denote the resulting embedded graph by $G'$. The edges in $E(G') \cap E(G)$ are called *original edges*, and edges in $E(G') \backslash E(G)$ are called *new edges*. Observe that this operation does not change distances in the graph. (In Figure 3, for the sake of readability we have omitted this first step, and assume that $G = G'$. This is no problem since this doubling step is only used to guarantee degree upper bounds for the resulting tree $T^*$.)

Now construct $\mathcal{R}_{G'}$, the radial graph of $G'$. Let $T_S$ be a BFS spanning tree of $\mathcal{R}_{G'}$, rooted at a vertex $r \in V(\mathcal{R}_{G'}) \cap V(G)$ of eccentricity $d$. Choose $T^*$ to be a spanning tree of the dual graph of $\mathcal{R}_{G'}$, such that for all edges $e^* \in E(T^*)$, the corresponding primal edge $e$ is not in $T_S$. Note that it is possible to construct such a spanning tree, since $T_S$ contains no cycles.

If the genus $g$ of the given map is zero, then at this point every edge of $\mathcal{R}_{G'}$ appears either in $T_S$, or its dual edge appears in $T^*$. However, if $g > 0$ then there are some edges for which neither property holds. Add all of these edges to $T_S$, to obtain the graph $T_S^+$. This completes the first stage of the construction of a surface split decomposition. For the proofs below we switch to a topological viewpoint, and assume that all of the given graphs are embedded in $\mathbb{S}_g$ with a 2-cell embedding, as discussed in Section 2.

We remark that there is a trivial bijection between the faces of $\mathcal{R}_{G'}$ and the edges of $G'$, since every face of $\mathcal{R}_{G'}$ contains exactly one edge of $G'$. Secondly, there is a trivial bijection between the vertices of $T^*$ and the faces of $\mathcal{R}_{G'}$, since $T^*$ is a subgraph of the dual graph of $\mathcal{R}_{G'}$. Combining these bijections yields a bijection from $V(T^*)$ to $E(G')$. In the remainder of this section, we refer to these bijections when speaking of 'the set of edges of $G'$ or faces of $\mathcal{R}_{G'}$ that *corresponds to* a subset of $V(T^*)$', or 'the set of edges of $G'$ that *corresponds to* a subset of the faces of $\mathcal{R}_{G'}$'. The *subgraph of $G'$ that corresponds to a set $T_1 \subseteq V(T^*)$* is the subgraph of $G'$ induced by the edges that correspond to $T_1$.

For a set $\mathcal{F}$ of faces of $\mathcal{R}_{G'}$, we define the open set $R(\mathcal{F}) \subseteq \mathbb{S}_g$ as follows: Informally, $R(\mathcal{F})$ is obtained by starting with $\mathcal{F}$ and 'connecting' all pairs of faces that share an edge. To be precise, to obtain $R(\mathcal{F})$ we first delete all edges and vertices of $\mathcal{R}_{G'}$ for which *all* incident faces are in $\mathcal{F}$. Call the resulting graph $\mathcal{R}'_{G'}$. Now $R(\mathcal{F})$ is the union of faces of $\mathcal{R}'_{G'}$ that contain points of faces in $\mathcal{F}$. Observe that if the set $\mathcal{F}$ corresponds to a connected subgraph of the dual graph of $\mathcal{R}_{G'}$, then $R(\mathcal{F})$ is connected, and therefore a region. For a set $T_1 \subseteq V(T^*)$, we denote by $R(T_1)$ the open set $R(\mathcal{F})$, where $\mathcal{F}$ is the set of faces of $\mathcal{R}_{G'}$ that corresponds to $T_1$. The following property now follows from the definitions and observations above.

**Lemma 12** *For an edge $e_T \in E(T^*)$, let $T_1$ and $T_2$ be the vertex sets of the two components of $T^* - e_T$, and let $G'_1$ and $G'_2$ be the subgraphs of $G'$ that correspond to $T_1$ and $T_2$ respectively. For every such edge $e_T$, the sets $R(T_1)$ and $R(T_2)$ are disjoint regions, such that for $i = 1, 2$, all vertices and edges of $G_i$ are drawn entirely in the closure of $R_i$.*

The above lemma shows that $T^*$ is already close to a surface split decomposition of $G'$, and therefore of $G$. The fact that it is not actually a branch decomposition yet (since $T^*$ is not necessarily ternary and since non-leaves may correspond to edges of $G$) will be addressed later. First we give a bound on what will be the width of the resulting branch decomposition, in Lemma 14. For this we need the following simple fact.

**Proposition 13** *Let $g$ be the genus of the map $G, \pi$. For $T_S^+$ and $T_S$ as constructed above, it holds that $|E(T_S^+)| - |E(T_S)| = 2g$.*

*Proof:* Denote $x = |E(T_S^+)| - |E(T_S)|$, and denote by $n$, $m$ and $f$ the number of vertices, edges and faces respectively of $\mathcal{R}_{G'}$. By Euler's formula, $2 - 2g = n - m + f$. (Recall that $G$, and therefore $\mathcal{R}_{G'}$, can be 2-cell embedded in $\mathbb{S}_g$.) The number of edges and arcs of $T_S^+$ is $n - 1 + x$, since $x$ edges were added to a spanning tree of $\mathcal{R}_{G'}$. The number of edges of $T^*$ is $f - 1$, since it is a spanning tree of the dual graph of $\mathcal{R}_{G'}$. For every edge $e$ of $\mathcal{R}_{G'}$, either $e$ appears in $T_S^+$ or its dual is part of $T^*$. Therefore, $m = (n-1+x) + (f-1)$. This yields $2 - 2g = n - m + f = 2 - x$, and thus $x = 2g$. $\square$

The following definitions are used in the proof of Lemma 14. An *Euler tour* is a walk that contains every edge of a graph exactly once. It is well-known that every connected graph in which all vertex degrees are even admits an Euler tour. A directed edge from vertex $u$ to $v$ is called an *arc*, and denoted by $(u, v)$. An *orientation* of an undirected (multi-)graph is obtained by replacing every edge $uv$ either by a arc $(u, v)$, or an arc $(v, u)$. In a *partial orientation*, only a subset of the edges is replaced by arcs. This results in a *mixed graph*, which contains both edges and arcs. The *in-degree* of a vertex $v$ in a mixed graph is the number of arcs $(u, v)$ (so incident edges are not counted). A *diwalk of length $k$* in a mixed (multi-)graph $G$ is a sequence $v_0, e_1, v_1, e_2, \ldots, v_k$ where $v_i \in V(G)$, and $e_i \in E(G)$ is an arc of $G$ from $v_{i-1}$ to $v_i$. It is a *dicycle* if $v_0, \ldots, v_{k-1}$ are distinct, but $v_0 = v_k$.

**Lemma 14** *For an edge $e_T \in E(T^*)$, let $T_1$ be the vertex set of one of the components of $T^* - e_T$, and let $G_1'$ be the subgraph of $G'$ that corresponds to $T_1$. For every such edge $e_T$, there are at most $(2g+1)(4d+3)/2$ vertices of $G_1'$ that lie on the boundary of $R(T_1)$.*

*Proof:* Consider the set $\mathcal{F}$ of faces of $\mathcal{R}_{G'}$ that corresponds to $T_1$. The boundary of the region $R(T_1)$ is formed by vertices and edges of $\mathcal{R}_{G'}$ that are incident both with faces from $\mathcal{F}$ and faces not in $\mathcal{F}$. Denote by $V_B \subseteq V(\mathcal{R}_{G'})$ and $E_B \subseteq E(\mathcal{R}_{G'})$ these sets. Thus $(V_B, E_B)$ is a subgraph of $\mathcal{R}_{G'}$, which we denote by $B$. We give an upper bound for $|E_B|$, which will then yield the stated upper bound on the number of vertices in $V(G') \cap V_B$.

Clearly all vertices in $V_B$ have even degree in $B$. Therefore we can construct a collection $W_1, \ldots, W_c$ of Euler tours, one for every component of $B$. We show that the length of such an Euler tour $W_i$ is bounded by $x(4d+3)$, where $x$ is the number edges in $W_i$ that are not part of $T_S$. Edges of $B$ that are not part of $T_S$ will be called *special edges*. There are two kinds of special edges: firstly, there is a single edge $e_T^P \in E(\mathcal{R}_{G'})$ that is the primal edge of $e_T \in E(T^*)$. This is the only edge of $B$ that is not part of $T_S^+$. Hence all other special edges are edges in $E(T_S^+) \setminus E(T_S)$. Therefore there are at most $2g + 1$ special edges in total in $B$ (Proposition 13).

Since $T_S$ is a BFS tree of $\mathcal{R}_{G'}$ rooted at a vertex $r \in V(G')$ of eccentricity $d$ (with respect to $G'$), the height of $T_S$ is at most $2d + 1$: this holds because every edge $uv$ in a walk in $G'$ can be replaced by a walk of length two in $\mathcal{R}_{G'}$ from $u$ to $v$. In addition, one extra edge may be necessary when considering the distance to a face vertex of $\mathcal{R}_{G'}$. (That is, a vertex that is not part of $G'$.)

Now we consider the following partial orientation of $B$: we orient the $T_S$-edges away from the root $r$. To be precise, we start with an orientation of $T_S$ such that the root has in-degree 0 and all other vertices have in-degree 1. All edges of $B$ that are in $T_S$ are oriented this way, and all other edges (special edges) remain undirected. We conclude that this partial orientation of $B$ satisfies the following properties.

- the maximum length of a diwalk in $B$ is $2d + 1$.

- $B$ contains no dicycles.
- $B$ contains no vertices with in-degree greater than 1.

It follows that every Euler tour $W_i$ contains at least one special edge, otherwise there would be a dicycle or a vertex with in-degree at least 2 in $B$. Let $W_i = v_0, \ldots, v_{\ell-1}, v_0$, and assume w.l.o.g. that $v_0 v_1$ is a special edge. A *segment* of $W_i$ is a maximal subwalk $W' = v_x, v_{x+1}, \ldots, v_y$ such that $v_x v_{x+1}$ is a special edge and $W'$ contains no other special edges. Hence the edges of $W_i$ can be partitioned into $x$ segments, where $x$ is the number of special edges in $W_i$. By the above properties, the length of a segment is bounded by $4d + 3$: in the extremal case, $W'$ consists of a special edge, a dipath of length $2d + 1$ oriented in the opposite direction of $W'$, and a dipath of length $2d + 1$ oriented in the same direction as $W'$. (Reversing the direction again would yield a vertex of in-degree at least 2.) As we observed above, the total number of special edges in $B$ is bounded by $2g + 1$. It follows that $|E_B| \leq (2g + 1)(4d + 3)$.

Since $B$ is a subgraph of $\mathcal{R}_{G'}$ and $\mathcal{R}_{G'}$ has a bipartition determined by the face vertices and the original vertices of $\mathcal{R}_{G'}$ (recall that original vertices are vertices in $V(\mathcal{R}_{G'}) \cap V(G')$), every edge in $E_B$ is incident with exactly one vertex of $G'$. Every vertex of $G'$ that lies on the boundary of $R(T_1)$ is incident with at least two edges from $E_B$. It follows that the number of those vertices is at most $|E_B|/2 \leq (2g + 1)(4d + 3)/2$. □

At this point, $T^*$ is already close to a low-width surface split decomposition of $G$, as shown by the Lemmas 12 and 14. However, $T^*$ is not yet a branch decomposition since it is not yet necessarily a ternary tree, and it does not hold that a vertex of $T^*$ is a leaf if and only if it corresponds to an original edge of $G$. We will now make some simple changes to guarantee these properties. We will use the following observation on the maximum degrees of $T^*$. Recall that original edges of $G'$ are edges that are also in $G$.

**Proposition 15** *A vertex $v_d \in V(T^*)$ has degree at most 2 if it corresponds to an original edge of $G'$, and degree at most 3 otherwise.*

*Proof:* First we prove that every vertex $v_d$ of $T^*$ that corresponds to an original edge $e = uv$ of $G'$ has degree at most 2. Because new edges have been added on either side of $e$, $e$ is incident with two 2-faces in $G'$. Hence in the radial graph $\mathcal{R}_{G'}$ of $G'$, the face $f$ that corresponds to $e$ is incident with the vertices $u$, $v$ and two vertices $x$ and $y$, with $d(x) = d(y) = 2$. Therefore, the spanning tree $T_S$ of $\mathcal{R}_{G'}$ contains at least one edge that is both incident with $x$ and with the face $f$. The same holds for $y$. It follows that the spanning tree $T^*$ of the dual graph of $\mathcal{R}_{G'}$ contains at most two edges incident with $v_d$.

Similarly, it follows that every vertex $v_d$ of $T^*$ that corresponds to a new edge $e$ in $G'$ has degree at most 3: the new edge $e$ is incident to at least one 2-face in $G'$. So at least one of the four edges of the face $f$ of $\mathcal{R}_{G'}$ that corresponds to $e$ is included in $T_S$. The claim follows. □

**Second stage of the construction** Using Proposition 15, we can modify $T^*$ as follows to obtain a new tree $T$ and mapping $\mu$ from $L(T)$ to $E(G)$: Firstly, for every vertex $v_d \in V(T^*)$ that corresponds to a new edge $e \in E(G')$, with $d(v_d) \leq 2$: delete $v_d$ if $d(v_d) = 1$, and suppress $v_d$ if $d(v_d) = 2$. Hence the remaining vertices of $T^*$ that correspond to new edges have degree 3. Subsequently, for every vertex $v_d \in V(T^*)$ that corresponds to an original edge $e \in E(G')$: if $d(v_d) = 1$, then set $\mu(v_d) = e$. If $d(v_d) = 2$, then introduce a new vertex $u$ and an edge $uv_d$, and mark $\mu(u) = e$. By Proposition 15 this concludes all possible cases.

*Proof of* Theorem 11: We prove that the tuple $(T, \mu)$ constructed above is a surface split decomposition of $G$, of width $(2g + 1)(4d + 3)/2$. From the operations above it immediately follows that $T$ is a ternary tree and $\mu$ is a bijection from the leaves of $T$ to the original edges of $G$, so $(T, \mu)$ is a branch decomposition of $G$. Consider an edge $e_T \in E(T)$, and the two tree components $T_1$ and $T_2$ of $T - e$. For $i = 1, 2$, let $G_i$ be the subgraph of $G$ (not of $G'$) induced by the edge set $\mu(L(T) \cap V(T_i))$.

First consider the case that one of the trees, say $T_1$, consists of a single vertex. Then $G_1$ consists of a single edge $e_G$. In this case it is trivial to define the disjoint regions $R_1$ and $R_2$ such that $e_G$ lies in the closure of $R_1$ and all other edges lie in the closure of $R_2$. Clearly $|\text{mid}(e_T)| \leq 2$ in this case.

Now suppose both $T_1$ and $T_2$ are non-trivial. Then by the construction of $T$, $e_T$ is also an edge of $T^*$, or $e_T$ is the result of suppressing a sequence of adjacent degree 2 vertices in $T^*$, that all correspond to new edges of $G'$. In either case, we easily select a corresponding edge $e^* \in E(T^*)$ that separates the vertices of $T^*$ into components $T_1^*$ and $T_2^*$, such that for $i = 1, 2$, $\mu(L(T) \cap V(T_i))$ is a subset of the edges of $G'$ that corresponds to $T_i^*$. Hence Lemma 12 shows that the desired regions $R_1$ and $R_2$ exist (recall that $G$ is a subgraph of $G'$). Furthermore, Lemma 14 shows that at most $(2g + 1)(4d + 3)/2$ vertices of $G$ lie on the boundary of these regions (recall that $V(G') = V(G)$). Hence $|\text{mid}(e_T)| \leq (2g + 1)(4d + 3)/2$.

Using the given combinatorial embedding $\pi$ of $G$, it can be shown using standard arguments and data structures that all graphs used during the construction of $(T, \mu)$ can be constructed in linear time (these are $G'$, $\mathcal{R}_{G'}$, $T_S$, $T^*$, $T_S^+$ and $T$ itself), together with their combinatorial embeddings and with data structures that keep track of the bijections between the various vertex/face/edge sets. □

## 6 Summary of the algorithm for bounded genus graphs

In this section we show how to combine the ingredients of the previous sections to obtain our main results; we give an algorithm for counting the number of subgraphs isomorphic to $P$ in a graph $G$ of bounded genus, and an algorithm for listing all of them. Without loss of generality, we may assume that $G$ is connected. We first use a relatively standard argument related to decomposing embedded graphs into layers that admit low width branch decompositions (Lemma 16). Eppstein used this argument for Subgraph Isomorphism in planar graphs [14], and in bounded genus graphs [15]. It was also used by Dorn [8].

Throughout this section we use the following notations. We choose an arbitrary vertex $r \in V(G)$, and construct a BFS tree $T_{\text{BFS}}$ rooted at $r$. Let $d$ be the eccentricity of $r$. For $i = 0, \ldots, d$, define $L_i \subseteq V(G)$ to be the set of vertices at distance $i$ of $r$, which are the vertices at height $i$ in $T_{\text{BFS}}$. We call this set a *layer*. Let $G_i^j = G[L_i \cup L_{i+1} \cup \ldots \cup L_j]$.

**Lemma 16** *Let $G, \pi$ be a map of genus at most $g$, with a vertex $r$ of eccentricity $d$, and subgraphs $G_i^j$ as defined above. For $i, j$ with $0 \leq i \leq j \leq d$, a surface split decomposition of $G_i^j$ of width at most $(2g + 1)(4j - 4i + 7)/2$ can be constructed in time $O(n')$, where $n' = |V(G_i^j)|$.*

*Proof:* For every $i$, let the graph $H_i^j$ be constructed from $G_i^j$ by adding one vertex $v$, and edges from $v$ to every vertex in $L_i$. This graph has genus at most $g$ as well, since it can be obtained from the graph $G[L_0 \cup L_1 \cup \ldots \cup L_j]$ by

- first deleting all edges that are incident with *at least one* vertex in $L_0 \cup \ldots L_{i-1}$ and that are not in $T_{\text{BFS}}$, and

- subsequently contracting all remaining edges with *both* end vertices in $L_0 \cup \ldots \cup L_{i-1}$.

Observe that contracting a set of edges that does not contain a cycle can be done while maintaining an embedding in $\mathbb{S}_g$. Clearly the vertex $v$ has eccentricity $j - i + 1$ in $H_i^j$, so by Theorem 11, a surface split decomposition of $H_i^j$ of width at most $(2g + 1)(4(j - i + 1) + 3)/2$ can be constructed in linear time. This easily yields a surface split decomposition of $G_i^j$ of width at most $(2g + 1)(4j - 4i + 7)/2$. □

We now present a novel algorithm for counting the number of $P$-isomorphs in $G$, even for the case where $P$ is disconnected. Let $c$ be the number of components of the pattern graph $P$, which are denoted by $P^1, \ldots, P^c$. For $S \subseteq \{1, \ldots, c\}$, denote by $P^S$ the subgraph of $P$ induced by the components with labels in $S$. Formally, $P^S = P[\cup_{i \in S} V(P^i)]$. For every $G_i^j$, we consider the following coloring, which uses the color set $\{1, \ldots, j-i+1\}$: the vertices of layer $L_x$ are all colored with color $x-i+1$. For $0 \leq i \leq j \leq d$ with $j-i < k$ and $S \subseteq \{1, \ldots, c\}$, we define $\text{DPC}_i^j(S)$ to be the number of *colorful* subgraphs of $G_i^j$ that are isomorphic to $P^S$. In other words, this is the number of $P^S$-isomorphs that use all layers of $G_i^j$.

**Proposition 17** *Let $g$ be a constant. Let $G, \pi$ be a simple map of genus at most $g$, let $P$ be a graph on $k$ vertices with $c$ components, and let $DPC_i^j(S)$ and $G_i^j$ be as defined above. For given $i, j$ with $0 \leq i \leq j \leq d$ and $j - i < k$, and given set $S \subseteq \{1, \ldots, c\}$, $DPC_i^j(S)$ can be computed in time $2^{O(k)} \cdot O(n')$, where $n' = |V(G_i^j)|$.*

*Proof:* By Lemma 16, we can find a surface split decomposition of $G'$ of width $w \in O(g(j-i)) \subseteq O(k)$ in time $O(n')$ (recall that $g$ is viewed as constant). The layer-based coloring for $G_i^j$ uses $q \leq j - i + 1 \leq k$ colors. Hence by Theorem 10, we can compute the number of colorful subgraphs of $G_i^j$ that are isomorphic to $P^S$ in time $2^{O(k)} \cdot O(n')$. □

For $j \in \{0, \ldots, d\}$, we define $\text{DPT}^j(S)$ to be the number of subgraphs of $G_0^j$ that are isomorphic to $P^S$. Note that these are not required to be colorful. In particular, if $S = \emptyset$, then we define $\text{DPT}^j(S) = 1$. To avoid the discussion of trivial cases, we simply define $\text{DPC}_i^j(S) = 0$ if $i < 0$, and $\text{DPT}^j(S) = 0$ if $j < 0$.

**Lemma 18** *For every $j \in \{0, \ldots, d\}$ and $S \subseteq \{1, \ldots, c\}$:*

$$DPT^j(S) = DPT^{j-1}(S) + \sum DPT^{j-x-1}(S_1) \cdot DPC_{j-x+1}^j(S_2),$$

*where the summation is over all partitions of $S$ into $S_1$ and $S_2$ with $S_2 \neq \emptyset$, and all $x \in \{1, \ldots, k\}$.*

*Proof:* We prove that every subgraph $G'$ of $G$ that is isomorphic to $P^S$ is counted exactly once. Consider such a subgraph $G'$. If $G'$ contains no vertices of layer $L_j$, then it counts towards $\text{DPT}^{j-1}(S)$ as well, and therefore it is counted by the first term. In the summation term, since $\text{DPC}_{j-x+1}^j(S)$ counts only colorful graphs and $x \geq 1$, only subgraphs that contain vertices of layer $L_j$ are considered. Hence in this case, $G'$ is counted exactly once.

Now suppose that $G'$ contains a vertex of layer $L_j$. Then it is not a subgraph of $G_0^{j-1}$, so it is not counted by the first term. Let $x$ be the maximum number such that $G'$ contains vertices of all layers $L_j, L_{j-1}, \ldots, L_{j-x+1}$. Since $P^S$ has at most $k$ vertices, we have $x \in \{1, \ldots, k\}$. Let $S_2 \subseteq S$ be the indices of components of $G'$ that are subgraphs of $G_{j-x+1}^j$, and let $S_1 \subseteq S$ be the indices of components of $G'$ that are subgraphs of $G_0^{j-x-1}$. By choice of $x$, $L_{j-x}$ contains no vertices of $G'$, so every component of $G'$ falls in one of the two categories, and therefore $S_1, S_2$ is a partition of $S$. Now we have deduced the unique combination of $x$, $S_1$ and $S_2$ such that $G'$ consists of a colorful subgraph of $G_{j-x+1}^j$ isomorphic to $P^{S_2}$, and a subgraph of $G_0^{j-x-1}$ that is isomorphic to $P^{S_1}$. Hence in this case, $G'$ is also counted exactly once, namely by the corresponding term in the summation. □

**Theorem 19** *Let $g$ be a constant. Let $G, \pi$ be a simple map on $n$ vertices of genus at most $g$, and let $P$ be a graph on $k$ vertices with $c$ components. In time $3^c \cdot 2^{O(k)} \cdot O(n) \subseteq 2^{O(k)} \cdot O(n)$, the number of subgraphs of $G$ that are isomorphic to $P$ can be computed.*

*Proof:* In the *first stage* of the algorithm, we compute $\mathrm{DPC}_i^j(S)$ for every $i, j$ with $0 \leq i \leq j \leq d$ and $j - i < k$, and every $S \subseteq \{1, \ldots, c\}$. There are $2^c$ choices of $S$, so computing $\mathrm{DPC}_i^j$ for one particular choice of $i$ and $j$ and all possible choices of $S$ takes time $2^c \cdot 2^{O(k)} \cdot O(n')$, where $n' = |V(G_i^j)|$ (Proposition 17). Since every vertex of $G$ appears in $G_i^j$ for at most $O(k^2)$ choices of $i$ and $j$, this yields a total complexity of $2^c \cdot 2^{O(k)} \cdot k^2 \cdot O(n) \subseteq 2^c \cdot 2^{O(k)} \cdot O(n)$ for the first stage.

In the *second stage* of the algorithm, we use the recursion from Lemma 18 for computing $\mathrm{DPT}^j(S)$ for every $j \in \{0, \ldots, d\}$ and $S \subseteq \{1, \ldots, c\}$. For a given $j$, it takes time $3^c \cdot O(k)$ to compute this for every $S$: there are $3^c$ partitions of $\{1, \ldots, c\}$ into $S_1$, $S_2$ and $\overline{S}$ (the complement of $S$), and $k$ choices of $x$, and therefore the total number of terms in all expressions (for given $j$) is bounded by $3^c \cdot k$. Computing this for every $j$ (in increasing order) therefore takes time $3^c \cdot O(k) \cdot O(d) \subseteq 3^c \cdot O(k) \cdot O(n) \subseteq 2^{O(k)} \cdot O(n)$.

Finally, the algorithm returns the value of $\mathrm{DPT}^d(\{1, \ldots, c\})$, which is the total number of subgraphs of $G_0^d = G$ that are isomorphic to $P$. □

Now we show how to extend the above counting algorithm to an algorithm for listing all isomorphs. For a connected graph $P$ on $k$ vertices, Eppstein [14] gave an algorithm for listing all $P$-isomorphs in a planar graph $G$ on $n$ vertices, in time $2^{O(k \log k)} \cdot O(n) + k^{O(1)} \cdot m$, where $m$ is the number of $P$-isomorphs. He asked whether this could be extended to the case where $P$ is disconnected. We answer this question, by extending the result to the case where $G$ may have bounded genus and $P$ may be disconnected. Furthermore, we improve the complexity. The proof of our result follows the general method sketched by Eppstein [14, Lemma 3], which uses three stages. We will call these the *counting stage*, *backtracking stage* and *generation stage*. We apply this to our dynamic programming algorithm from Theorem 19. In addition, we fill in some essential details of the analysis, related to data structures, which is necessary to obtain the claimed complexity bound.

**Theorem 20** *Let $g$ be a constant. Let $G, \pi$ be a simple map on $n$ vertices of genus at most $g$, and let $P$ be a graph on $k$ vertices with $c$ components. In time $3^c \cdot 2^{O(k)} \cdot O(n) + mk^{O(1)}$, all subgraphs of $G$ that are isomorphic to $P$ can be generated, where $m$ is the number of such subgraphs.*

*Proof:* The entire listing algorithm has three stages: a *counting stage*, a *backtracking stage*, and a *generation stage*. For the *counting stage*, we first run the counting algorithm described in the proof of Theorem 19. Recall that this algorithm first computes the values $\mathrm{DPC}_i^j(S)$ for every $S \subseteq \{1, \ldots, c\}$ and $0 \leq i \leq j \leq d$ with $j - i < k$ using the algorithm from Section 3, and then computes the values $\mathrm{DPT}^j(S)$ for every $S \subseteq \{1, \ldots, c\}$ and $j \in \{0, \ldots, d\}$ using the recursion from Lemma 18.

For the *backtracking stage*, we compute for every $\mathrm{DPT}^j(S)$ and $\mathrm{DPC}_i^j(S)$ whether they contribute to the total number of isomorphs, or *contribute* for short. This notion is defined recursively as follows, and denoted by $0/1$ variables $\mathrm{CT}^j(S)$ and $\mathrm{CC}_i^j(S)$, respectively.

- We initialize $\mathrm{CT}^j(S) = 0$ for every $j$ and $S$, and $\mathrm{CC}_i^j(S) = 0$ for every $i, j, S$.

- Recall that the total number of subgraphs of $G$ that are isomorphic to $P$ is $\mathrm{DPT}^d(\{1, \ldots, c\})$. Therefore, we set $\mathrm{CT}^d(\{1, \ldots, c\}) = 1$.

- Every entry $\mathrm{DPT}^j(S)$, in particular $\mathrm{DPT}^d(\{1, \ldots, c\})$, is computed by a summation of $\mathrm{DPT}^{j-1}(S)$ and a number of terms $\mathrm{DPT}^{j-x-1}(S_1) \cdot \mathrm{DPC}_{j-x+1}^j(S_2)$ (Lemma 18). For every entry $\mathrm{DPT}^j(S)$ that contributes to the total (i.e. that has $\mathrm{CT}^j(S) = 1$), and every non-zero term $\mathrm{DPT}^{j-x-1}(S_1) \cdot \mathrm{DPC}_{j-x+1}^j(S_2)$ in its sum, we set $\mathrm{CT}^{j-x-1}(S_1) = 1$ and $\mathrm{CC}_{j-x+1}^j(S_2) = 1$. If $\mathrm{DPT}^{j-1}(S) \neq 0$, then we set $\mathrm{CT}^{j-1}(S) = 1$ as well.

In the next stage we will generate lists of graphs for every $j$, $S$ with $\text{CT}^j(S) = 1$, and every $i$, $j$, $S$ with $\text{CC}^j_i(S) = 1$. We have now set these values such that every graph that is generated is also part of at least one $P$-isomorph in $G$. It is easily seen that this backtracking stage has the same complexity as the counting algorithm, which is $3^c \cdot O(k) \cdot O(n)$. Here we only consider the part of the counting algorithm where the entries $\text{DPT}^j(S)$ are computed, which is called the 'second stage' in the proof of Theorem 19.

Now we consider the last stage, the *generation stage*. Suppose first that for every $i$, $j$ and $S$, where $\text{DPC}^j_i(S) = x$ and $\text{CC}^j_i(S) = 1$, we have a list $LC^j_i(S)$ that contains the $x$ different colorful subgraphs of $G^j_i$ that are isomorphic to $P^S$. (How to obtain this list will be discussed later.) For the generation stage, we essentially run the counting algorithm again. However, every time that we consider an entry $\text{DPT}^j(S) = y$ with $\text{CT}^j(S) = 1$, instead of recomputing $\text{DPT}^j(S)$, we construct a list $LT^j(S)$ that contains the $y$ corresponding subgraphs of $G^j_0$. We construct the list $LT^j(S)$ corresponding to $\text{DPT}^j(S)$ as follows. First, we take a copy of the list $LT^{j-1}(S)$. Secondly, for every non-zero term $\text{DPT}^{j-x-1}(S_1) \cdot \text{DPC}^j_{j-x+1}(S_2)$ in its sum (Lemma 18), we combine every graph in the list $LT^{j-x-1}(S_1)$ with every graph in the list $LC^j_{j-x+1}(S_2)$. (These lists have been computed before since we have set $\text{CT}^{j-x-1}(S_1) = 1$ and $\text{CC}^j_{j-x+1}(S_2) = 1$ in the backtracking stage.) The resulting list of graphs is appended to the list $LT^j(S)$. By doing this for every $\text{DPT}^j(S)$ with $\text{CT}^j(S) = 1$, this procedure ends with the construction of a list $LT^d(\{1,\ldots,c\})$, which is the desired list of all subgraphs of $G$ that are isomorphic to $P$.

However, to obtain the stated complexity bound, with an additive term $mk^{O(1)}$ instead of $nmk^{O(1)}$, we have to be careful here. Consider for example the case that $G$ contains $m'$ subgraphs that are isomorphic to $P$, and are contained in layer $L_1$ (i.e. are subgraphs of $G[L_i]$). Let $\mathcal{C} = \{1,\ldots,c\}$. Then the list $LT^1(\mathcal{C})$ has length $m'$. If we *copy* this list every time, when constructing the lists $LT^j(\mathcal{C})$ for $j = 2,\ldots,d$, then we use $\Omega(dm')$ copy operations. Both $m'$ and $d$ may be much larger than $k$, so this cannot be bounded by the stated complexity $2^{O(k)} \cdot O(n) + mk^{O(1)}$. On the other hand, if for the computation of $LT^j(S)$ we start with the existing list $LT^{j-1}(S)$, and append entries, then the original list $LT^{j-1}(S)$ is not available anymore for later computations.

Therefore, we initialize the list $LT^j(S)$ in constant time with a *pointer* to the list $LT^{j-1}(S)$, and later interpret this as the list $LT^{j-1}(S)$ itself. If the list $LT^{j-1}(S)$ itself consists of a single pointer to another list $LT^{j'}(S)$, and contains no further entries, then instead we let the pointer for $LT^j(S)$ point to the list $LT^{j'}(S)$ as well. This ensures that finding an actual entry from a list takes only constant time; there is at most one pointer lookup for every actual entry that we find.

Similarly, for every term $\text{DPT}^{j-x-1}(S_1) \cdot \text{DPC}^j_{j-x+1}(S_2)$ where $S_1 = \emptyset$ (and thus $\text{DPT}^{j-x-1}(S_1) = 1$), the resulting list of graphs is the same as $LC^j_{j-x+1}(S_2)$. In this case, we append a pointer to $LC^j_{j-x+1}(S_2)$ to the list $LT^j(S)$ in a similar way.

Based on this implementation, we can now prove the desired complexity bound. For every $j$ and $S$, every operation involved in constructing the list $LT^j(S)$ that does not consist of combining non-empty graphs from different list takes constant time. Therefore, the complexity of all of these operations can be bounded by $3^c \cdot 2^{O(k)} \cdot O(n)$ again (with the same argument as in the proof of Theorem 19). Because of the backtracking stage, we only construct graphs that will be part of $P$-isomorphs of $G$. Therefore, every operation that combines (non-empty) graphs from a list $LT^{j-x-1}(S_1)$ with graphs from a list $LC^j_{j-x+1}(S_2)$ can be attributed to the construction of a $P$-isomorph. Every such operation increases the number of components in the resulting graph (we take the disjoint union of non-empty graphs in $G^{j-x-1}_0$ with graphs in $G^j_{j-x+1}$). Hence for every $P$-isomorph there are at most $c \le k$ such operations, which all take time polynomial in $k$. Hence the complexity of all of these operations can be bounded by $mk^{O(1)}$. This proves that the complexity of this part of the algorithm is bounded by $3^c \cdot 2^{O(k)} \cdot$

$O(n) + mk^{O(1)}$.

It remains to consider the problem of computing a list of graphs corresponding to entries $\mathrm{DPC}_i^j(S)$ with $0 \le j - i < k$ and $\mathrm{CC}_i^j(S) = 1$. That is, we have to construct a list of all colorful subgraphs of $G_i^j$ that are isomorphic to $P^S$. This can be done analogously, based on the counting algorithm from Section 3. The algorithm again consists of a counting stage, backtracking stage, and a generation stage. Let $(T, \mu)$ be the constructed surface split decomposition of $G_i^j$. For every $e \in E(T)$ and every entry $(T, \gamma, A, \eta) \in \mathcal{T}_e$, we first compute whether the entry contributes to the total number of colorful $P^S$-isomorphs. If so, we construct a list of $\eta$ subgraphs of $G_e$ that correspond to this entry. Similar to the above argument, we use pointers. This way, every operation that results in a list of graphs that is the same as a previous list takes constant time, and can be attributed to the counting part of the complexity function. (This happens for instance when combining the empty graph from a list $\mathcal{T}_f$ with graphs from $\mathcal{T}_g$, where $f$ and $g$ are the children of $e$.) Every operation where graphs in lists are modified is attributed to one of the final colorful $P^S$ isomorphs.

With a similar argument as before we can guarantee that only polynomially many operations are attributed to every such isomorph: When we combine entries such that modifying a list is necessary, in every case the number of vertices plus edges increases, or the number of $\gamma$-images decreases. This yields the stated complexity. □

## 7 Extensions and improvements

**Non-orientable surfaces** Although we restricted ourselves to orientable surfaces for ease of presentation, we remark that our results hold for non-orientable surfaces as well. We shortly mention the facts about non-orientable surfaces that have to be used for adapting our proofs to this case. For $c \ge 1$, denote by $\mathbb{N}_c$ the surface obtained by adding $c$ *crosscaps* to $\mathbb{S}_0$ (see [18] for details). $\mathbb{N}_c$ is called the *non-orientable surface of genus c*. For every non-orientable surface $S$, there exists exactly one value $c \ge 1$ such that $S$ is homeomorphic to $\mathbb{N}_c$ [18, p.82]. For a graph $G$ with $n$ vertices and $m$ edges that is 2-cell embedded in $\mathbb{N}_c$ with $f$ faces, Euler's formula states that $n - m + f = 2 - c$. For graphs $G$ that are embedded in $\mathbb{N}_c$, the dual graph $G^*$ can be defined similarly as before, with one vertex for every face of $G$, and the same number of edges as $G$. (See [18, p.103] for details.) Similarly, the radial graph $\mathcal{R}_G$ of $G$ can be defined. With these ingredients, the constructions, bounds and proofs of Section 5 can easily be generalized to the non-orientable case.

With regard to the bound in Section 4, Bender and Canfield [3] considered non-orientable maps as well, and proved that there are at most $O(m^{5(c-2)/4} 12^m)$ simple rooted maps of non-orientable genus $c$ on $m$ edges. We remark that two non-orientable embeddings of a graph $G$ in $\mathbb{N}_c$ are called equivalent if facial walks are mapped to facial walks; we cannot define this based on clockwise edge orders now. Equivalence of non-orientable maps is defined similarly. Although Miller [17] treated only the orientable case explicitly, Grohe [16] analyzed a different isomorphism testing algorithm for both the orientable and non-orientable case, and proved that it decides isomorphism correctly in time $n^{O(g)}$ and $n^{O(c)}$, respectively. This shows how the results from Section 4 can be generalized.

In Section 6, the only observation related to embeddings was that contracting a set of edges that does not contain cycles can be done while maintaining an embedding in $\mathbb{S}_g$. Clearly, this holds for $\mathbb{N}_c$ as well. In the remaining sections, no information about the embedding was used. We conclude that our results hold for the non-orientable case as well (with different constants).

**Induced subgraphs** We observe that with a trivial modification, our algorithm can find and count *induced* subgraphs as well. Consider the dynamic programming algorithm from Section 3. For every edge $e \in E(T)$ of the branch decomposition, we now define the table $\mathcal{T}_e$ to be *induced*

$k$-correct if for every entry $(H, \gamma, A, \eta) \in \mathcal{T}_e$, there are exactly $\eta$ *induced* subgraphs $G'$ of $G_e$ with $|V(G')| \le k$ that are $(H, \gamma, A)$-subgraphs, and $\mathcal{T}_e$ contains no pairs of equivalent entries. Note that it should be an induced subgraph with respect to $G_e$, not necessarily with respect to the entire graph $G$. For leaf edges $e \in E(T)$, this property is easily guaranteed: in this case, $G_e$ is a graph that consists of a single edge $xy$. When initializing the table $\mathcal{T}_e$, we omit the subgraph that consists of two vertices and no edges, and include the other four subgraphs of $G_e$. This yields an induced $k$-correct table. Observe that the dynamic programming update rule from Section 3 preserves induced $k$-correctness, without further modifications. (However, we remark that a table may contain graphs that do not correspond to induced subgraphs of $G$; $G$ may contain an edge between two vertices of $G_e$ that is not part of $G_e$.) Hence Theorem 10 holds for Induced Subgraph Isomorphism as well. The algorithm from Theorem 19 does not have to be modified either: it combines induced subgraphs of $G_{j-x+1}^j$ with induced subgraphs of $G_0^{j-x-1}$. Since these graphs are separated by the layer $L_{j-x}$, the result is again an induced subgraph.

**Planar graphs** For bounding the table size in Section 4, we applied a bound by Bender and Canfield [3], which is expressed in the number of edges of the graph. In the case of planar graphs, we can apply a result by Tutte, based on the number of vertices of the graph, to obtain better constants. Tutte [24] showed that there are $O(9.5^n)$ equivalence classes of rooted triangulated maps on $n$ vertices. We can modify the encoding in the proof of Lemma 7 as follows: given an entry $(H, \gamma, A, \eta)$, we can embed $H$ such that all $\gamma$-images are incident with the outer face, introduce the vertex $u$ again, drawn in the outer face, and add edges from $u$ to every $\gamma$-image in $H$. Choose one of these edges to be the root, similar to before. Next, we add edges until a triangulation $H', \pi'$ is obtained. The edge labeling $\lambda$ is constructed by marking the added edges with 0, and the original edges with 1. This yields an encoding $H', \pi', (u, v), \lambda, \rho$ of the pair $H, \gamma$, where $\rho$ and $(u, v)$ are defined the same as before. Then $H', \pi'$ is a rooted triangulated map on at most $k+1$ vertices. Hence it has at most $3k-3$ edges. Applying Tutte's result [24] and assuming that only one color is used (which may be done for the connected case, see [14]), it follows that there are at most $2^w \cdot O(9.5^k) \cdot 2^{3k-3} = 2^w \cdot O(76^k)$ entries in the table. Although this constant is still very large, it is significantly better than the constants from the previous algorithm for planar graphs [8]. This is a strong indication that when implemented, our algorithm is likely to be much faster as well. We believe that for small values of $k$ and connected graphs $P$, our algorithm may in fact be fast enough in practice, contrary to what our rough asymptotic complexity bounds suggest. Developing and testing a good implementation would be a good topic for further research. The key to a fast implementation would be a faster dynamic programming update step, based on ordered tables (see [19, 7] for examples).

**Acknowledgement** The author would like to thank Frederic Dorn for the introduction to the subject and the many inspiring discussions.

# References

[1] J. ALBER, H. FERNAU, AND R. NIEDERMEIER, *Parameterized complexity: exponential speed-up for planar graph problems*, J. Algorithms, 52 (2004), pp. 26–56.

[2] N. ALON, R. YUSTER, AND U. ZWICK, *Color-coding*, J. ACM, 42 (1995), pp. 844–856.

[3] E. A. BENDER AND E. R. CANFIELD, *The asymptotic number of rooted maps on a surface*, J. Combin. Theory Ser. A, 43 (1986), pp. 244–257.

[4] E. D. DEMAINE, F. V. FOMIN, M. HAJIAGHAYI, AND D. M. THILIKOS, *Subexponential parameterized algorithms on bounded-genus graphs and $H$-minor-free graphs*, J. ACM, 52 (2005), pp. 866–893.

[5] E. D. DEMAINE AND M. HAJIAGHAYI, *The bidimensionality theory and its algorithmic applications*, Comput. J., 51 (2008), pp. 292–302.

[6] R. Diestel, *Graph theory*, vol. 173 of Graduate Texts in Mathematics, Springer, Heidelberg, fourth ed., 2010.

[7] F. Dorn, *Designing subexponential algorithms: problems, techniques and structures*, PhD thesis, University of Bergen, Norway, 2007.

[8] ———, *Planar subgraph isomorphism revisited*, in STACS 2010, vol. 5 of LIPIcs, Dagstuhl, Germany, 2010, Schloss Dagstuhl–Leibniz-Zentrum fuer Informatik, pp. 263–274.

[9] F. Dorn, F. V. Fomin, and D. M. Thilikos, *Fast subexponential algorithm for non-local problems on graphs of bounded genus*, in Algorithm theory—SWAT 2006, vol. 4059 of LNCS, Springer, Berlin, 2006, pp. 172–183.

[10] F. Dorn, E. Penninkx, H. L. Bodlaender, and F. V. Fomin, *Efficient exact algorithms on planar graphs: exploiting sphere cut branch decompositions*, in Algorithms—ESA 2005, vol. 3669 of LNCS, Springer, Berlin, 2005, pp. 95–106.

[11] ———, *Efficient exact algorithms on planar graphs: exploiting sphere cut decompositions*, Algorithmica, 58 (2010), pp. 790–810.

[12] R. G. Downey and M. R. Fellows, *Parameterized complexity*, Springer-Verlag, New York, 1999.

[13] D. Eppstein, *Subgraph isomorphism in planar graphs and related problems*, in SODA'95, New York, 1995, ACM, pp. 632–640.

[14] ———, *Subgraph isomorphism in planar graphs and related problems*, J. Graph Algorithms Appl., 3 (1999), pp. 1–27.

[15] ———, *Diameter and treewidth in minor-closed graph families*, Algorithmica, 27 (2000), pp. 275–291.

[16] M. Grohe, *Isomorphism testing for embeddable graphs through definability*, in STOC'00, New York, 2000, ACM, pp. 63–72.

[17] G. Miller, *Isomorphism testing for graphs of bounded genus*, in STOC'80, New York, 1980, ACM, pp. 225–235.

[18] B. Mohar and C. Thomassen, *Graphs on surfaces*, Johns Hopkins Studies in the Mathematical Sciences, Johns Hopkins University Press, Baltimore, MD, 2001.

[19] R. Niedermeier, *Invitation to fixed-parameter algorithms*, vol. 31 of Oxford Lecture Series in Mathematics and its Applications, Oxford University Press, Oxford, 2006.

[20] J. Plehn and B. Voigt, *Finding minimally weighted subgraphs*, in WG '90, vol. 484 of LNCS, Springer, Berlin, 1991, pp. 18–29.

[21] J. Rué, I. Sau, and D. Thilikos, *Dynamic programming for graphs on surfaces*, in ICALP '10, vol. 6198 of LNCS, Springer Berlin / Heidelberg, 2010, pp. 372–383.

[22] P. D. Seymour and R. Thomas, *Call routing and the ratcatcher*, Combinatorica, 14 (1994), pp. 217–241.

[23] H. Tamaki, *A linear time heuristic for the branch-decomposition of planar graphs*, in Algorithms - ESA 2003, vol. 2832 of LNCS, Springer Berlin / Heidelberg, 2003, pp. 765–775.

[24] W. T. Tutte, *A census of planar triangulations*, Canad. J. Math., 14 (1962), pp. 21–38.